\documentclass[12pt]{iopart}
\usepackage{graphicx} 
\usepackage[bookmarks,bookmarksnumbered=true]{hyperref}
\usepackage[section]{placeins}
\usepackage{algorithm}
\usepackage{algpseudocode}

\begin{document}

\title{Training and synchronizing oscillator networks with Equilibrium Propagation}

\author{Théophile Rageau$^1$ \& Julie Grollier$^1$}

\address{$^1$ Laboratoire Albert Fert, CNRS, Thales, Université Paris-Saclay, 1 avenue Augustin Fresnel,
91767 Palaiseau Cedex,
France}
\eads{\mailto{theophile.rageau@cnrs-thales.fr}, \mailto{julie.grollier@cnrs-thales.fr}}
\vspace{10pt}
\begin{indented}
\item[]April 2025
\end{indented}

\begin{abstract}
Oscillator networks represent a promising technology for unconventional computing and artificial intelligence. Thus far, these systems have primarily been demonstrated in small-scale implementations, such as Ising Machines for solving combinatorial problems and associative memories for image recognition, typically trained without state-of-the-art gradient-based algorithms. Scaling up oscillator-based systems requires advanced gradient-based training methods that also ensure robustness against frequency dispersion between individual oscillators.
Here, we demonstrate through simulations that the Equilibrium Propagation algorithm enables effective gradient-based training of oscillator networks, facilitating synchronization even when initial oscillator frequencies are significantly dispersed. We specifically investigate two oscillator models: purely phase-coupled oscillators and oscillators coupled via both amplitude and phase interactions.
Our results show that these oscillator networks can scale successfully to standard image recognition benchmarks, such as achieving nearly 98\% test accuracy on the MNIST dataset, despite noise introduced by imperfect synchronization. This work thus paves the way for practical hardware implementations of large-scale oscillator networks, such as those based on spintronic devices.

\end{abstract}

\vspace{2pc}

\section*{Introduction}
Unconventional computing technologies are becoming essential for the new generation of artificial intelligence systems. Among these, oscillator networks are one of the most promising \cite{todri2024computing}. Various hardware implementations show that this technology can already be used to solve complex tasks such as combinatorial problems \cite{wang2019oim}\cite{moy20221}\cite{bazzi2024optimizing}\cite{english2022ising}\cite{chou2019analog}\cite{delacour2023mixed}\cite{graber2024integrated}, visual pattern recognition \cite{holzel2013neural}\cite{nikonov2020convolution}\cite{choi2025hardware} as well as to recognize spoken vowels with high precision \cite{torrejon2017neuromorphic}\cite{romera2018vowel}. From a biological perspective, neurons in the brain behave like nonlinear oscillators \cite{buzsaki2006rhythms} and can exhibit synchronization phenomena \cite{strogatz1993coupled}\cite{lehnertz2009synchronization}\cite{malagarriga2015synchronization} in a way comparable to coupled oscillators.
Networks of coupled oscillators have been applied to image recognition tasks through associative memory methods \cite{hoppensteadt1999oscillatory}.
The principle of associative memory networks is to assign a local minimum of network energy to each image. This can be achieved by setting the local couplings between oscillators or using one single time-dependent coupling that interacts differently with each oscillator.
The most common model of coupled oscillators used in associative memory networks is the Kuramoto model \cite{kuramoto1975self}. In the last decades, several approaches have been developed to extend this method to different kinds of oscillator networks (such as spin-torque \cite{popescu2018simulation} or ring oscillators \cite{rudner2024design}) and to increase its performances \cite{follmann2014phase}.
However, one remaining limitation of this approach is that it scales poorly to larger and more complex sets of images \cite{holzel2013neural}, especially compared to artificial neural networks with gradient-based approaches. Associative memory combined with gradient-based approaches seems to increase overall performance \cite{rudner2024design}.

To train such a physical system (i.e., to adjust the weights of the connections between each pair of oscillator neurons), the standard Backpropagation algorithm \cite{rumelhart1986learning} (or its variant for temporal tasks \cite{werbos1990backpropagation}) faces several challenges. The non-locality of its learning rule (derived by the chain rule) and the different types of neural computations involved (forward and backward passes implement different computations) make its implementation at the hardware level very difficult \cite{kuninti2021backpropagation}.
In this context, several energy-based learning algorithms have been developed to train analog models (see \cite{scellier2024energy} for a comparison on standard databases). In particular, the Equilibrium Propagation algorithm \cite{scellier2017equilibrium} is a learning framework that has the potential to solve the issues of the Backpropagation algorithm for on-chip training. It is a gradient-based learning framework, implementable at the hardware level, for a neural network with dynamic neurons. The algorithm alternates between relaxing the network to its minimum energy configuration and updating the network weights in the direction that minimizes a cost function. Compared to Backpropagation, this algorithm provides a local learning rule to update the coupling weights, and its inference and learning phases involve the same type of computations. To date, physical implementations of the Equilibrium algorithm have been made with resistive networks  \cite{yi2023activity}\cite{dillavou2024machine} and Ising Machines deployed on the D-Wave platform \cite{laydevant2024training}.
Developments around this algorithm are still ongoing and have made it possible to get closer and closer to Backpropagation's performance on large databases \cite{nest2024towards}.

Coupled-phase oscillator network could be an alternative hardware to implement Equilibrium Propagation, as shown in some recent simulation results \cite{wang2024training}. However, in the proposed framework, its implementation assumes that oscillators remain synchronized during training, which is rarely the case in physical oscillatory systems. For example, spin-torque nano-oscillators in a network can reach frequency dispersions values 5\% \cite{jenkins2024impact}, which makes a full synchronization difficult to achieve in practice \cite{tsunegi2018scaling}.

In this work, we adapt the Equilibrium Propagation algorithm to a network of coupled-phase oscillators with different natural frequencies. In particular, we show that Equilibrium Propagation enables training oscillatory systems while synchronizing oscillators. To demonstrate this property, we focus mainly on the Kuramoto model \cite{kuramoto1975self}, which is a standard framework to describe the dynamics and synchronization of phase-coupled oscillators \cite{rodrigues2016kuramoto}. Then, we investigate the universal nonlinear oscillator model \cite{slavin2009nonlinear} that captures the coupled phase-amplitude dynamics of a wide range of nonlinear oscillators. 
Specifically, the contributions of the present work are the following:

\begin{itemize}
    \item We achieve a test accuracy of $97.77\%$ on the MNIST database, representing the best known results using a network of fully synchronized Kuramoto phase-coupled oscillators trained with Equilibrium Propagation.
    \item We demonstrate analytically and with simulations that Equilibrium Propagation enables synchronization of a large population of oscillators during training.
    \item We show that the learning procedure of Equilibrium Propagation holds even with oscillator frequency dispersions of around $5$-$10\%$. In that case, we are able to achieve test accuracies of $96$-$97\%$ on the MNIST database.
    \item We adapt the framework provided by the Equilibrium Propagation algorithm to the universal nonlinear oscillator model that considers coupled amplitude-phase oscillators. We show that this model can solve the MNIST database with a test accuracy of $96.85\%$ using parameters representative of hardware spintronic systems.
\end{itemize}

\section{General Kuramoto model}\label{s:fully_synchronized_Kuramoto_model}

\Figure{\label{f:Kuramoto_net}
(A): Model of an oscillator in a network. The oscillator is represented by a vector in polar coordinates. In the Kuramoto model, the radius (amplitude of the oscillator) is fixed. The angle corresponds to its phase rotating at its frequency. The oscillator is driven by sources through unidirectional synapses (dotted line). It interacts with other oscillators using bidirectional synapses (solid line).
(B): Diagram of a network of phase-coupled oscillators that follows the oscillator model in Figure A. An input image is fed to the network. Its pixels are encoded in the phases of the input sources (in blue) that drive the hidden oscillators using unidirectional synapses (dotted lines). The hidden and output oscillators mutually influence their dynamics through bidirectional synapses (solid lines). Output oscillators encode the predicted label of the input image.
\includegraphics[scale=0.7]{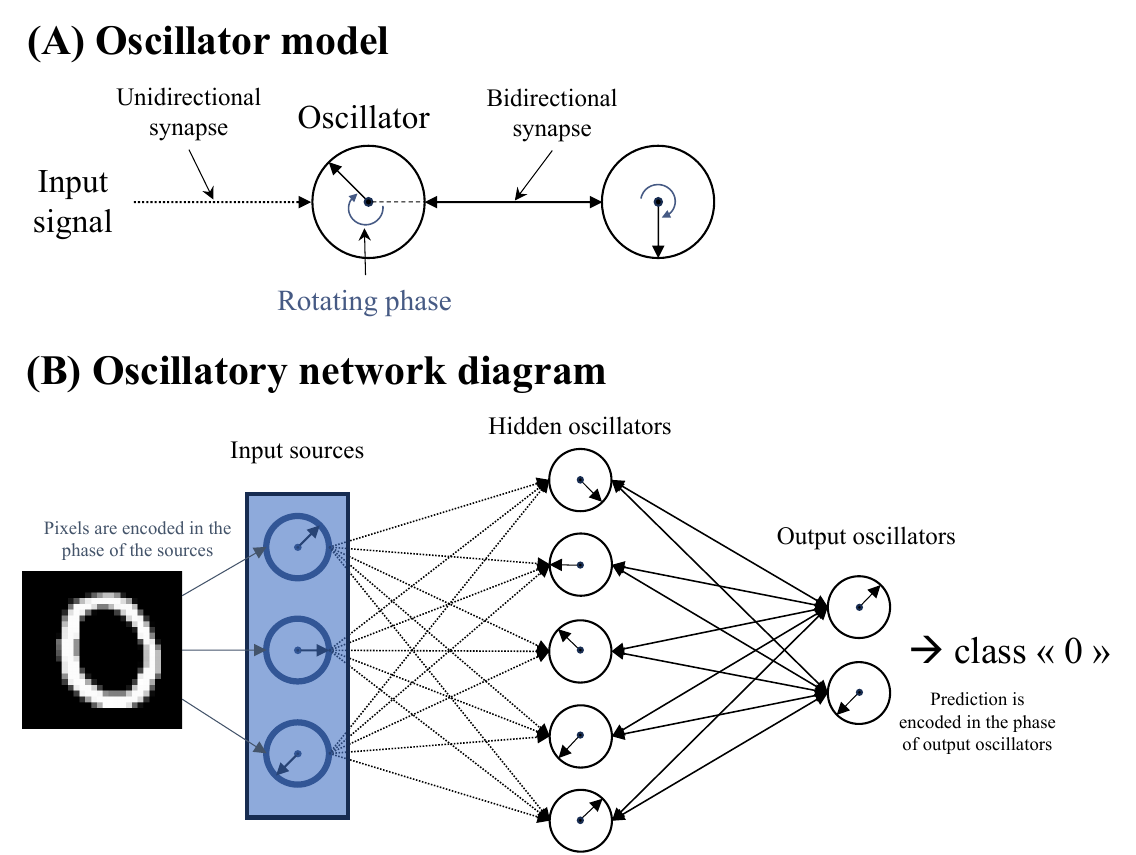}}

\subsection{Dynamics of phase-coupled oscillators} \label{s:nonsynch_Kuramoto}

We consider a network of coupled-phase oscillators governed by the Kuramoto model as represented in Figure \ref{f:Kuramoto_net} (see Figure (\ref{f:Kuramoto_net}A) for the oscillator model and Figure (\ref{f:Kuramoto_net}B) for the network), where the phases of the connected oscillators continuously interact and influence each other. In this model, oscillators are described solely by their phases, with amplitudes assumed to be constant.
Mathematically, the phase dynamics is given, in the stationary frame, by the system of coupled equations:
\begin{eqnarray} \label{e:nonsynch_Kuramoto}
    \frac{\rmd\phi_j}{\rmd t} + \omega_j = \sum_{k=1}^K \Omega_{j,k}\sin(\phi_k - \phi_j), \hspace{0.3cm} \forall j \in \lshad1,K\rshad
\end{eqnarray}

\noindent
where:

$\phi = (\phi_j)_{(1 \leq j \leq K)}$: oscillator phases

$\omega = (\omega_j)_{(1 \leq j \leq K)}$: oscillators' natural frequencies

$\Omega = (\Omega_{j,k})_{(1 \leq j,k \leq K)}$: coupling matrix

In this equation, the oscillator phases operate at their natural frequencies $\omega_j$, which may differ from one oscillator to another. The phase dynamics is influenced by coupling terms $\Omega_{j,k}$ which strengthen or reduce phase interactions between one oscillator and its neighbors.

\subsection{Oscillator synchronization in a network} \label{s:oscillator_synchro}
In some situations, oscillators in a network may synchronize partially or completely, in which case their phases start to operate at the same frequency, called synchronization frequency.
More precisely, a set of coupled oscillators defines a locking range that depends on the strength of the interactions. If the natural frequency of an oscillator is inside this range, its instantaneous frequency will be locked to the synchronization frequency.

\Figure{\label{f:phase_evolution}
Phase evolution of hidden layer oscillators for two different frequency dispersions (FD). The phases are observed in the rotating frame at frequency $\omega_0$ (frequency of the input sources). Figure (A) shows the phase evolution of oscillators with no frequency dispersion, and (B) with a frequency dispersion of 0.5\%. The vertical dashed black line on the left side shows the transition from a transient regime during which phases are reconfigured to a synchronization regime in which phases differences are fixed. This synchronization regime does not necessarily occur in the case of a network with frequency dispersion, as illustrated in Figure (B) in which some phases do not settle to a stable configuration.
\includegraphics[scale=0.5]{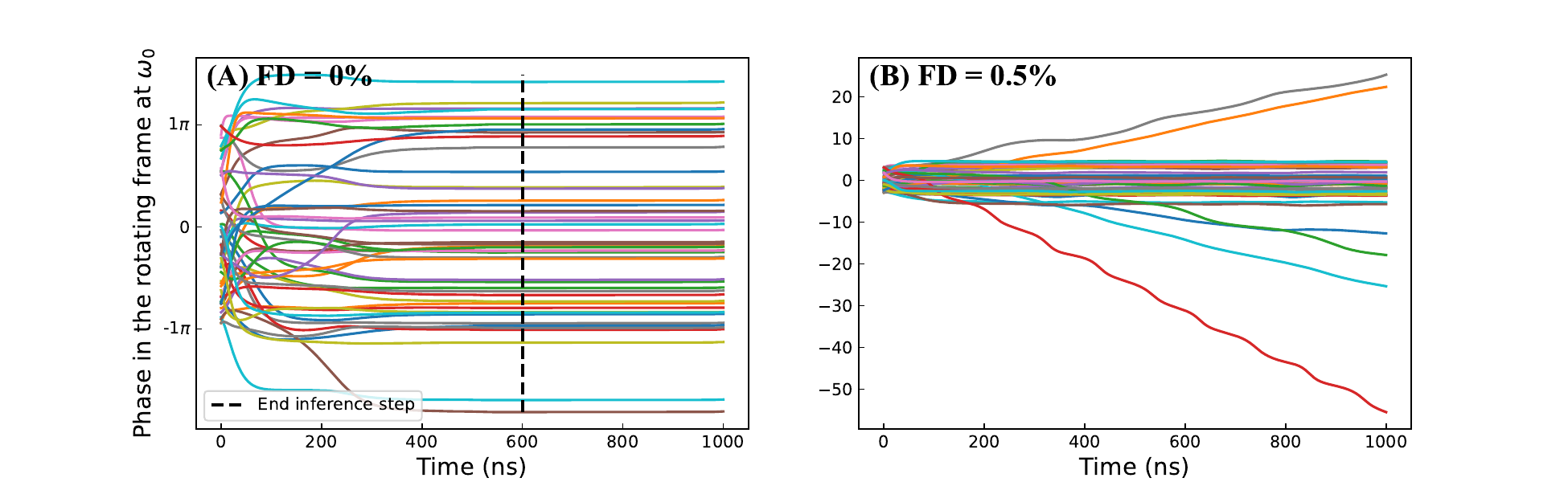}}

We illustrate the synchronization phenomena in Figure \ref{f:phase_evolution}. We simulate a fully-connected layered network (see \ref{s:network_architecture} for more details on the network architecture) of coupled oscillators with 64 input sources, which emit at the same frequency $\omega_0$, connected to 50 hidden oscillators, themselves connected to 10 output oscillators. The natural frequencies of the hidden and output oscillators are sampled in a Gaussian distribution around the common frequency of the sources $\omega_0$. We vary the frequency dispersion by tuning the standard deviation of the distribution. The coupling weights between the hidden oscillators, the input sources, and the output oscillators are randomly set (see Section \ref{s:network_architecture} for the weight initialization).
Figure \ref{f:phase_evolution} shows the evolution of the oscillator phases in the hidden layer for two networks with different frequency dispersions (FD). Phases are plotted in the rotating frame at the frequency of the sources $\omega_0$.
The first network (Figure \ref{f:phase_evolution}A) has no frequency dispersion, whereas the other (Figure \ref{f:phase_evolution}B) has a frequency dispersion of 0.5\%. In Section \ref{s:simulation_details} we provide more details on the simulation framework (physical considerations and technical details).

In Figure \ref{f:phase_evolution}A, we observe that phases are first reconfigured and then, they all stabilize to a synchronized state at the frequency of the sources with fixed phase differences. The vertical dashed black line shows the transition between the reconfiguration regime and the stationary regime (from 800 nanoseconds).
In Figure \ref{f:phase_evolution}B, not all phases stabilize after some time. Some of them diverge. These oscillators are not synchronized at the source frequency. Their natural frequency is outside the locking range of frequencies. However, most oscillators synchronize and act as in Figure \ref{f:phase_evolution}A but the non-synchronized oscillators disturb the stable configuration.

\section{Training an oscillatory network with Equilibrium Propagation}

\subsection{Adaptation of the Kuramoto model to the framework of Equilibrium Propagation}\label{s:Kuramoto_EqProp}

To train a network of phase-coupled oscillators with the Equilibrium Propagation algorithm, we need to describe its dynamics from an energy point of view, in order to deduce the learning rule of the model parameters, i.e. the coupling matrix $\Omega = (\Omega_{i,j})_{i,j \in \lshad 1, K \rshad}$. The internal energy of the system $E_\phi$ derives from the dynamic equation \ref{e:nonsynch_Kuramoto}. The time evolution of the system minimizes the energy:
\begin{eqnarray}
    \frac{\partial E_\phi}{\partial \phi_j} = -\frac{\rmd\phi_j}{\rmd t}
\end{eqnarray}
Hence, the phase energy in the stationary frame is:
\begin{eqnarray}\label{e:phase_energy}
    E_{\phi} := \sum_{j=1}^K \omega_j\phi_j - \sum_{j=1}^K \sum_{k=1}^K \Omega_{j,k}\cos(\phi_k - \phi_j)
\end{eqnarray}

Oscillators in the last layer of the network, called output oscillators (see Figure \ref{f:Kuramoto_net}A for example), are used as read-outs for the predictions made by the system. They aim to replicate the target that corresponds to a given input. The distance between the output units and the targets is measured by a cost function $\mathcal{L}$. The key idea of Equilibrium Propagation is to make this cost function act like a reconfigurable potential in the energy, that tends to drive the phases of the output oscillators towards a target. To control the strength of this potential, we introduce a scalar parameter $\beta$ called the 'nudging factor' and define the total energy function (as described in \cite{scellier2017equilibrium}):
\begin{eqnarray} \label{e:reconfigurable_energy}
    F_{\phi}(\phi, \Omega, \psi_{\rm{target}}, \beta) := E_{\phi}(\phi, \Omega, \psi_{\rm{target}}) + \beta\mathcal{L}(\phi, \psi_{\rm{target}})
\end{eqnarray}
With

$\phi = (\phi_j)_{(1 \leq j \leq K)}$: oscillator phases

$\Omega = (\Omega_{j,k})_{(1 \leq j,k \leq K)}$: learning parameters

$\psi_{\rm{target}} = (\psi_{\rm{target},o})_{(1 \leq o \leq O)}$: source phases that encode the target

$\beta$: nudge factor

$\mathcal{L}$: loss (or cost) function

Different loss functions $\mathcal{L}$ can be considered as energy potentials. In \cite{scellier2017equilibrium} and \cite{laydevant2024training}, the authors proposed using the standard mean squared error with a network of continuous or binary Ising spins. For a similar system in \cite{laborieux2021scaling}, the cross-entropy loss function is introduced. Otherwise in \cite{wang2024training}, a network of synchronized oscillators is trained with a variant of the cross-entropy loss function adapted to the system. All these loss functions are designed to discriminate between good and bad predictions of the output neurons.
In our approach, we choose a loss function intrinsically related to phase-coupled oscillator dynamics. We couple the output oscillators to sources (oscillators with fixed phases) that encode the target. The coupling between both sets of oscillators is set to the 'nudging factor' $\beta$:
\begin{eqnarray*}
    \beta\mathcal{L}(\phi, \psi_{\rm{target}}) := \sum_{o=1}^O \beta\cos(\psi_{\rm{target},o} - \phi_o)
\end{eqnarray*}
Depending on the ratio between the internal coupling matrix $\Omega$ and the external coupling $\beta$, the network relaxation will minimize the total energy $F_\phi$ by giving different importance to the two terms.

The Equilibrium Propagation algorithm has been extensively detailed in \cite{scellier2017equilibrium}\cite{ernoult2019updates}\cite{laborieux2021scaling}. We give here a quick reminder on the global learning procedure focusing on important details for coupled oscillators (see \ref{s:pseudo_code_EP} for the pseudo-code).

For each pair of input and target, the algorithm consists of two phases, referred to as the `free' and `nudge' phases.
In the `free' phase, we set the 'nudging factor' $\beta$ to 0 and let the network relax to a stable configuration $\phi^{0}$. The phase configuration achieved corresponds to a local minimum of the internal energy $E_\phi$. In the `nudge' phase, we set $\beta$ to a non-zero value and then let the network converge to another stable configuration $\phi^{\beta}$ which corresponds to a local minimum of the total energy $F_\phi$. This second local energy minimum is slightly different from the one obtained in the `free' phase. It has been influenced by the external energy potential $\beta\mathcal{L}$ in the total energy $F_\phi$, which means that the values of the output neurons are closer to the actual targets in $\phi^\beta$ than in $\phi^0$.
Then, the deviation between both stable configurations provides the gradient update of the learning parameter $\Omega$:
\begin{eqnarray} \label{e:general_learning_rule}
\Delta \Omega \propto -\frac{1}{\beta} \left( \frac{\partial F_{\phi}}{\partial\Omega}(\phi^{\beta}, \Omega, \beta) - \frac{\partial F_{\phi}}{\partial\Omega}(\phi^{0}, \Omega, 0) \right)
\end{eqnarray}
Finally, we write the new values of the learning parameters and iterate the whole procedure for another pair of input and target.
In our oscillatory system, we derive the learning rule of the coupling parameters $\Omega_{j,k}$ from the general equation \ref{e:general_learning_rule}.
If we denote respectively $\phi^{0} = (\phi^{0}_1, ..., \phi^{0}_K)$ and $\phi^{\beta} = (\phi^{\beta}_1, ..., \phi^{\beta}_K)$ the stable configurations of the phases after respectively the `free' and `nudge' phases, the following formula gives us the gradient updates for the parameter $\Omega_{j,k}$:
\begin{eqnarray} \label{e:synch_gradients}
    \Delta \Omega_{j,k} \propto \frac{1}{\beta} \left ( \cos(\phi^\beta_k - \phi^\beta_j) - \cos(\phi^0_k - \phi^0_j) \right )
\end{eqnarray}
As described in \cite{scellier2017equilibrium}, for vanishing values of $\beta$, these formulas approximate the gradient of the loss function $\nabla_\theta \mathcal{L}$. Hence, following this direction with a gradient-descent algorithm (such as stochastic-gradient-descent \cite{ruder2017overview} or Adam algorithm \cite{kingma2014adam}), we should be able to reach a minimum of $\mathcal{L}$ and solve the desired Machine Learning task.

\subsection{Learning in a network of non-synchronized oscillators}\label{s:learning_nonsynch}
A key assumption of Equilibrium Propagation is that it requires a system with converging units. It assumes that at the end of the relaxation (`free' and `nudge' phases), the network settles in a stable configuration.
However, as we saw in Section \ref{s:oscillator_synchro}, this requirement might not be fulfilled in an oscillatory network composed of oscillators with different natural frequencies.
If the oscillators are synchronized at the same frequency, we can consider their phase in the rotating frame at the synchronization frequency and they can be seen as converging units, as shown in Figure \ref{f:phase_evolution}A.
If all oscillators are not synchronized, it is still possible to consider their phase in the same rotating frame, but the phases of non-synchronized oscillators will diverge, as shown in Figure \ref{f:phase_evolution}B.

In practice, we observed that a subgroup of oscillators still synchronizes at the frequency of the sources. This can be explained by the fact that the frequencies of these oscillators are within the locking range. For these oscillators, it is therefore possible to define an internal coupling energy in the rotating frame of the sources (as given in equation \ref{e:phase_energy}) and thus to deduce the learning rules.

For connections including a non-synchronized oscillator, phases do not rotate at the same frequency, and the requirements of Equilibrium Propagation are not met. However, it turns out that the learning rule used in the case of synchronized oscillators allows the initially non-synchronized oscillators to synchronize during training. In fact, when updating a connection between two oscillators, including one not synchronized, each cosine in the learning rule \ref{e:synch_gradients} will be a random value between -1 and 1 (because the phases do not evolve at the same frequency). Then, the update value $\Delta\Omega_{j,k}$ will also be a random number in the interval $[-\frac{2}{\beta}, \frac{2}{\beta}]$. This implies that the weights of the connections will evolve randomly if both connected oscillators are not synchronized at the same frequency. If, during this random walk, the weight of a connection encounters the synchronization region of the network, the oscillators will become synchronized, and the weight of the connection will then evolve according to the learning rule for synchronized oscillators. More details and examples on synchronizing oscillators during training can be found in \ref{s:nonsynch_coupling_simulations}.

Hence, the coupling changes during training modify both the synchronization state and the expressive power of the network. Therefore, there exists a trade-off between finding coupling weights that enable the synchronization of a large population of oscillators (to have the largest number of available neurons) and finding weights that enable minimizing the cost function (to solve the task).

\subsection{Evolution of synchronization and accuracy during training}\label{s:evolution_synch_acc}

\Figure{\label{f:synchronization_accuracy_CV2_CV5} 
Evolution of accuracy and proportion of synchronized oscillators during training. We trained two networks of Kuramoto oscillators with a respective initial frequency dispersion of 2\% (A) and 5\% (B) on a small image database. In each figure, the blue curve shows the average proportion of synchronized oscillators over the training period (with standard deviations). The red curve shows in parallel the evolution of the accuracy of the model's learned data during training.\\[0.5cm]
\includegraphics[scale=0.45]{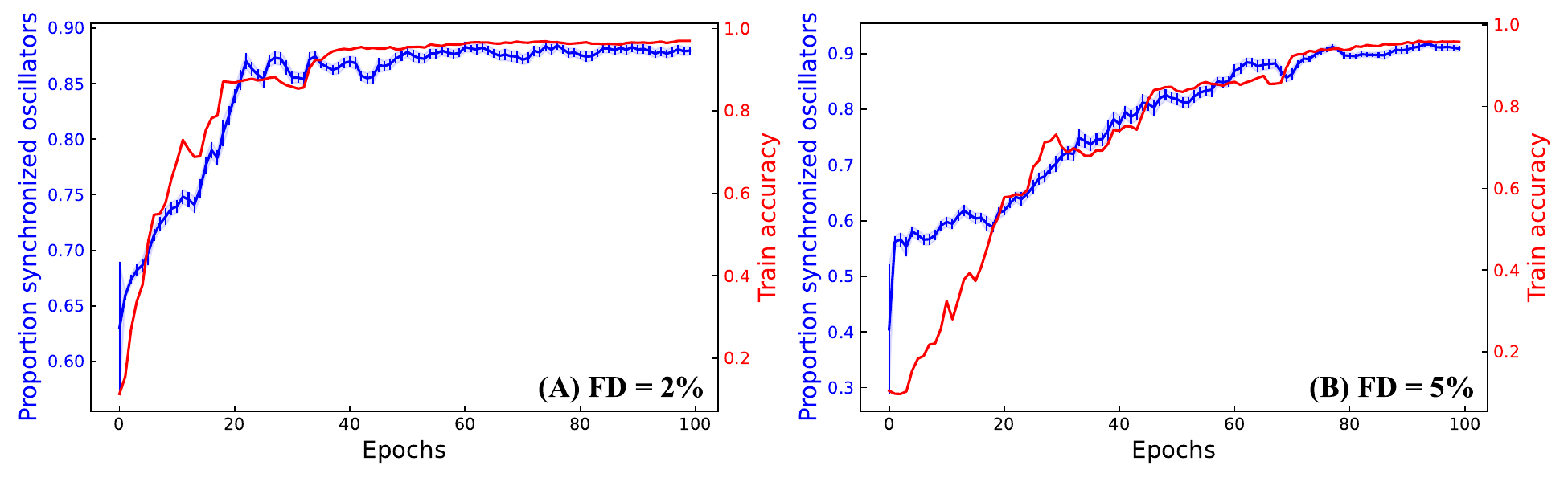}}

To illustrate how oscillator synchronization occurs during training, we focus on a small neural network (one hidden layer network with 50 hidden oscillators)trained with Equilibrium Propagation on a rescaled version of the MNIST database called Digits (\url{https://scikit-learn.org/stable/modules/generated/sklearn.datasets.load_digits.html}).
Figure \ref{f:synchronization_accuracy_CV2_CV5} shows in parallel the evolution of accuracy and the proportion of synchronized oscillators during training for two networks of Kuramoto oscillators with respectively a frequency dispersion of 2\% (Figure \ref{f:synchronization_accuracy_CV2_CV5}A) and 5\% (Figure \ref{f:synchronization_accuracy_CV2_CV5}B). For each input image, we count the number of oscillators that synchronize at the frequency of the sources in the hidden layer of the network during the `free' phase. Then, we averaged over all the training images for each epoch.

Several points can be highlighted from this figure.
First, we observe that both networks can learn with different frequency dispersions and synchronization is also increasing during training.
Both accuracy (red) and synchronization (blue) curves evolve in a very similar way. This indicates that oscillators globally manage to synchronize during training. As mentioned earlier, weight updates for non-synchronized oscillators are random, in contrast to the deterministic updates observed in synchronized ones. We find that these random updates tend to be significantly larger, allowing non-synchronized oscillators to undergo substantial frequency shifts that can eventually bring them into the synchronization range.
A demonstration of this phenomenon using statistical tools is given in Section \ref{s:nonsynch_coupling_analysis}.
A second observation from these curves is that, after a certain number of epochs, the system reaches a convergent state  corresponding to an optimal balance found between network synchronization and its capacity of learning.
Finally, for high-frequency dispersion (Figure \ref{f:synchronization_accuracy_CV2_CV5}B), there is a delay at the beginning of the training between the synchronization and accuracy curves. In fact, in the first few training epochs, the accuracy gets stuck to its initial value, whereas the synchronization increases. This is due to an initial intense reconfiguration of the network's couplings during the first training epochs. Indeed, at the beginning of  training, the coupling values are random, so they do not favor synchronization within the network (the proportion of synchronized oscillators after the first epoch is around 30\%). The iterative gradient updates of the couplings configure the network in a state that favors more oscillator synchronization (around 55\% after 3 epochs). Once this new state is obtained, phases can be arranged to process information, which helps the network learn.
This phenomenon appears very often while training a network with a high initial frequency dispersion (from 4\% to 5\%). For a smaller frequency dispersion (Figure \ref{f:synchronization_accuracy_CV2_CV5}A), it is smaller or is even not apparent.

\section{Performance on MNIST database}
In this section, we investigate the performance of oscillatory networks on the standard MNIST database \cite{lecun1998gradient} in three different configurations: fully-synchronized Kuramoto model, general Kuramoto model and coupled amplitude-phase model.

The MNIST database, composed of 70000 28x28 greyscale images of handwritten digits, from 0 to 9, including 60000 images for the train set and 10000 for the test set, is a  standard classification task for testing Machine Learning models.
Previous results have shown that a network of fully-synchronized coupled-phase oscillators can be trained with Equilibrium Propagation on a rescaled version of MNIST (8x8 images) comprising less than 2000 digit images \cite{wang2024training}.

\subsection{Fully-synchronized Kuramoto model}

The first configuration we investigate is a particular case of the general Kuramoto model described in \ref{s:nonsynch_Kuramoto}. Here, we assume that all oscillators share the same natural frequency $\omega_0$, which is also the frequency of the sources. Then, the oscillators are, by definition, synchronized at their mutual frequency $\omega_0$. This particular case removes all synchronization issues described in Section \ref{s:learning_nonsynch}.
To tackle the full MNIST database, we use a fully-connected-layered network with one hidden layer of 500 neurons (see section \ref{s:network_architecture} for a detailed description of the fully-connected architecture). We also add external bias sources at frequency $\omega_0$ to oscillators as proposed in \cite{wang2024training}. This increases the number of learning parameters in our system. In Section \ref{s:bias_impact}, we discuss their impact on the network performance. The total size of our network is 784-500-10 ($28\times28 = 784$ pixels as input, 500 hidden neurons, and 10 output classes).
We look for optimal values of the hyper-parameters. More details can be found in Section \ref{s:hyperparameters}.

The comparison of the results is summarized in Table \ref{t:comparison_Ising_Kuramoto}. Both columns on the left show benchmark results obtained by training a standard Ising model with Equilibrium Propagation, described by a different energy function from the one we use here. With this model, \cite{scellier2017equilibrium} obtains an accuracy of 97.5\% and \cite{ernoult2019updates} reaches 97.94\% by enhancing the precision used to simulate the dynamics of the system.
With our system, we reach an average test accuracy of 97.77\% \ensuremath{\pm} 0.08\% (mean over 5 runs) on the MNIST database as shown in the right column of Table \ref{t:comparison_Ising_Kuramoto}. Therefore, our model compares very well with benchmark results for an equivalent Ising network trained with Equilibrium Propagation.

\Table{\label{t:comparison_Ising_Kuramoto}Comparison of performance (test accuracy) for standard Ising networks and the oscillatory Kuramoto network on MNIST database.}
\br\\
\ns\ns\ns
\centre{2}{Ising networks}&\centre{1}{\textbf{Kuramoto network}}\\
\bs
\crule{2}&\crule{1}\\
\bs
Scellier \& Bengio (2017)& Ernoult \& al. (2019) & \textbf{Rageau \& Grollier (2025)}\\
\mr
$\sim 97.5\%$&$97.94 \pm 0.16\%$&$\mathbf{97.77 \pm 0.08\%}$\\
\br
\endTable

\subsection{General Kuramoto model}
In the previous section, we made the assumption that all oscillators have exactly the same natural frequency $\omega_0$. However, this is rarely achieved in real physical systems. This is mostly due to the fact that the natural frequency of a device strongly depends on its fabrication process, yielding variability between devices on the same substrate. For several types of oscillators such as spin-torque nano-oscillators, frequency dispersions in a network may reach values above 5\% \cite{jenkins2024impact}.

Here, we simulate a network of Kuramoto oscillators with different natural frequencies, sampled in a Gaussian distribution around the frequency of the sources $\omega_0$, and train it on the MNIST database. We vary the frequency dispersion (FD) by tuning the standard deviation of the Gaussian distribution. The frequency dispersion measures the rate of displacement compared to the mean frequency. The standard deviation is then equal to the frequency dispersion times the mean frequency.

\Figure{\label{f:accuracy_differentCV} Evolution of accuracy and `nudge factor' (in MHz) as a function of frequency dispersion around the mean frequency $\omega_0=4.2$GHz. For each value of the frequency dispersion, we trained a network on the MNIST database and recorded the best accuracy obtained on the test set. The blue curve shows the evolution of the accuracy as the frequency dispersion within the oscillator network increases. The red curve shows in parallel the evolution of the `nudge factor' we had to use to obtain these results.
\includegraphics[scale=0.7]{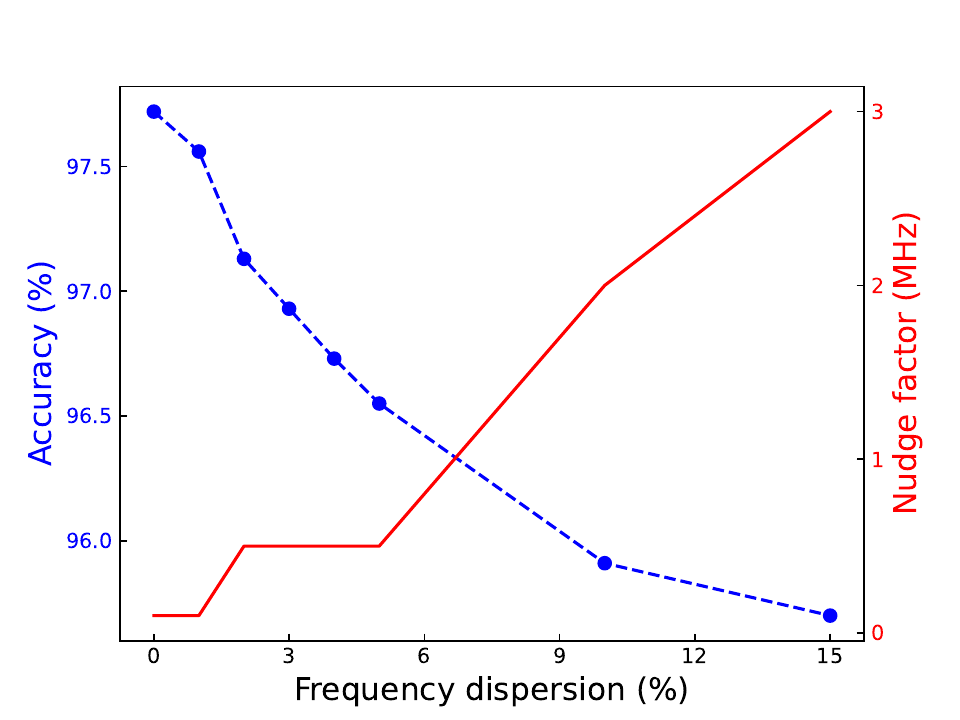}}

Figure \ref{f:accuracy_differentCV} shows the evolution of the accuracy on MNIST for each frequency dispersion. We observe that the performance of our model deteriorates as frequency dispersion increases. However, this reduction is moderate and we still achieve more than 95\% accuracy on the test set with high frequency dispersions (10\% to 15\%).
The key to maintaining high performance even with large dispersions is to properly choose the `nudge factor' $\beta$. In our simulations, we observed that for high frequency dispersions, small values of $\beta$ led to poor quality learning while increased values of the nudge factor lead to improved results. This phenomena can be explained by the internal dynamics of our network. Indeed, couplings between oscillators tend to synchronize them at a certain frequency of the sources $\omega_0$. However a few oscillators may rarely or never synchronize to the others if their natural frequency is too distant from the average frequency of the network. These few oscillators still influence the dynamics of the network and act like background noise on the output oscillator phases. Then, by using a small `nudging factor', the network cannot differentiate between an actual nudging term and noise. Thus, the greater the number of non-synchronized oscillators, the larger the required `nudging factor'.

These results show that training a network of Kuramoto phase-coupled oscillators with Equilibrium Propagation is robust to frequency variability. This opens up perspective for future hardware implementation, in which frequency variability can be a strong issue.

\subsection{Fully-synchronized coupled amplitude-phase model}

In this section, we look at a more general model of coupled oscillators. Whereas the Kuramoto model assumes fixed oscillator amplitudes, we consider here that the oscillators are instead characterized by an amplitude and a phase that operate in a coupled way. For this purpose, we use the universal nonlinear oscillator model with negative damping and nonlinear frequency shift proposed in \cite{slavin2009nonlinear}. This model is very general and can be used to describe different types of auto-oscillatory systems. 

Following the theory described in \cite{slavin2009nonlinear}, the dynamics of a system of $K$ coupled nonlinear oscillators is described by the following coupled equations for powers $p$ (amplitude squared) and phases $\phi$ (the derivation of both equations can be find in \ref{s:derivation_Slavin}):
\begin{eqnarray} \label{e:Slavin_dynamics}
\eqalign{
      & \frac{\rmd p_j}{\rmd t} + 2p_j\left( \Gamma_+(p_j) - \Gamma_-(p_j) \right) = 2\sum_{k=1}^K \Omega_{k,j}\sqrt{p_j p_k}\cos(\phi_k - \phi_j + \beta_{k,j})\\
      & \frac{\rmd\phi_j}{\rmd t} + \omega_j(p_j) = \sum_{k=1}^K \Omega_{k,j}\sqrt{\frac{p_k}{p_j}}\sin(\phi_k - \phi_j + \beta_{k,j})
}
\end{eqnarray}

\noindent
where:

$p = (p_j)_{(1 \leq j \leq K)}$: oscillator powers in the network

$\phi = (\phi_j)_{(1 \leq j \leq K)}$: oscillator phases in the network

$\omega_j(p) = \omega_{0,j} + Np$: frequency of oscillator $j$ 

$\Gamma_+(p) = \Gamma_G(1 + Qp)$: positive damping

$\Gamma_-(p) = \sigma I(1 - p)$: negative damping

$\Omega = (\Omega_{j,k})_{(1 \leq j,k \leq K)}$: coupling matrix

$\beta_{k,j}$: coupling phase between oscillators $k$ and $j$

In the phase equation, several terms are similar to those that appear in the Kuramoto dynamics (see equation \ref{e:nonsynch_Kuramoto}) such as pairwise couplings $\Omega_{i,j}$ and the frequency term $\omega_j$ (more information on this model can be found in \ref{s:nonlinear_frequency_shift}).

From this dynamics, we adapt this oscillator model to the Equilibrium Propagation framework, and derive the learning rule of the coupling matrix $\Omega$ (see derivations and results in \ref{s:derivation_Slavin}).

We tested the performance of this model on the MNIST database. To be able to compare the results with those obtained by the Kuramoto model, we use the same network architecture as in the previous cases (i.e. fully-connected network with one hidden layer of size 784-500-10 and external bias sources). We also assumed that oscillators and bias in the network all have the same natural frequency $\omega_0$.

With the coupled amplitude-phase model of oscillator, we were able to obtain $96.85 \pm 0.16\%$ accuracy on the MNIST database. More information on the physical considerations and technical details (optimization, computational power and simulations issues) are given in \ref{s:simulation_details}.

One particular observation we made during the fine-tuning process was that this kind of network never reached 100\% accuracy on the training set, which is usually the case when training deep learning models. Moreover, this phenomenon occurred regardless of the database considered (smaller databases such as 2D databases, or larger one as MNIST). For each trial, there was a gap of around 1\% between perfect precision and precision on the training set. Taking this into account, the result for the test set appears to be more consistent, with a generalization gap of around 2\% between the training set and the test set. We observed the same kind of generalization gap for the Kuramoto model.

The results obtained are about 1\% less than those obtained with the Kuramoto model. As the Kuramoto model directly derives from the nonlinear universal model, we might have expected similar accuracies for comparable network architectures and hyperparameters (see Table \ref{t:hyperparameters} in Appendix for a comparison).
We attribute the difference to the fact that we use an approximation of the energy function to derive the learning rules of the model that may reduce the precision of weight updates in the last stage of training (see \ref{s:derivation_Slavin} for more information).

\section*{Conclusion}
In this work, we have shown through simulations that a network of phase-coupled oscillators can be a candidate device for a hardware implementation of the Equilibrium Propagation algorithm.
We demonstrated that Equilibrium Propagation intrinsically enables synchronization in oscillatory networks through a random walk of weights connecting non-synchronized oscillators.
The results presented in Table \ref{t:comparison_Ising_Kuramoto} show that a network of phase-coupled Kuramoto oscillators is capable of obtaining results comparable to those of the state of the art with an Ising model of spins on the MNIST database.
We have also shown that training a network of this type is relatively robust to frequency variability, which is common in certain oscillator models.
Finally, we looked at the spintronic model of coupled oscillators with phase and amplitude and showed that it is possible to train it with a 1\% accuracy loss on the MNIST database.

\section*{Conflict of Interest Statement}
The authors declare that the research was conducted in the absence of any commercial or financial relationships that could be construed as a potential conflict of interest.

\section*{Author Contributions}
The project was designed by JG and TR. The code and numerical simulations were all carried out by TR. TR and JG wrote the manuscript.

\section*{Funding}
This work was funded by the European Research
Council advanced grant GrenaDyn (reference: 101020684) and  by a public grant overseen by the French
National Research Agency (ANR) as part of the ‘PEPR IA France 2030’
program (Emergences project ANR-23-PEIA-0002

\section*{Acknowledgments}
This project was supported with AI computing and storage resources by GENCI at IDRIS thanks to Grant 2023-AD010913993 on the supercomputer Jean Zay's V100 partition.

\section*{Supplemental Data}

The programming code utilized in this study will soon be available.

\section*{References}
\bibliographystyle{IEEEtran}
\bibliography{bibliography}

\begin{thebibliography}{10}
\providecommand{\url}[1]{#1}
\csname url@samestyle\endcsname
\providecommand{\newblock}{\relax}
\providecommand{\bibinfo}[2]{#2}
\providecommand{\BIBentrySTDinterwordspacing}{\spaceskip=0pt\relax}
\providecommand{\BIBentryALTinterwordstretchfactor}{4}
\providecommand{\BIBentryALTinterwordspacing}{\spaceskip=\fontdimen2\font plus
\BIBentryALTinterwordstretchfactor\fontdimen3\font minus \fontdimen4\font\relax}
\providecommand{\BIBforeignlanguage}[2]{{%
\expandafter\ifx\csname l@#1\endcsname\relax
\typeout{** WARNING: IEEEtran.bst: No hyphenation pattern has been}%
\typeout{** loaded for the language `#1'. Using the pattern for}%
\typeout{** the default language instead.}%
\else
\language=\csname l@#1\endcsname
\fi
#2}}
\providecommand{\BIBdecl}{\relax}
\BIBdecl

\bibitem{todri2024computing}
A.~Todri-Sanial, C.~Delacour, M.~Abernot, and F.~Sabo, ``Computing with oscillators from theoretical underpinnings to applications and demonstrators,'' \emph{Npj unconventional computing}, vol.~1, no.~1, pp. 1--16, 2024.

\bibitem{wang2019oim}
T.~Wang and J.~Roychowdhury, ``Oim: Oscillator-based ising machines for solving combinatorial optimisation problems,'' in \emph{Unconventional Computation and Natural Computation: 18th International Conference, UCNC 2019, Tokyo, Japan, June 3--7, 2019, Proceedings 18}.\hskip 1em plus 0.5em minus 0.4em\relax Springer, 2019, pp. 232--256.

\bibitem{moy20221}
W.~Moy, I.~Ahmed, P.-w. Chiu, J.~Moy, S.~S. Sapatnekar, and C.~H. Kim, ``A 1,968-node coupled ring oscillator circuit for combinatorial optimization problem solving,'' \emph{Nature Electronics}, vol.~5, no.~5, pp. 310--317, 2022.

\bibitem{bazzi2024optimizing}
A.~Bazzi, E.~Hardy, J.~Ballester, F.~Badets, and L.~Hutin, ``Optimizing with phases: design and application space assessment for networks of phase-locked ring oscillators,'' in \emph{2024 IEEE 6th International Conference on AI Circuits and Systems (AICAS)}.\hskip 1em plus 0.5em minus 0.4em\relax IEEE, 2024, pp. 602--606.

\bibitem{english2022ising}
L.~Q. English, A.~Zampetaki, K.~Kalinin, N.~Berloff, and P.~G. Kevrekidis, ``An ising machine based on networks of subharmonic electrical resonators,'' \emph{Communications Physics}, vol.~5, no.~1, p. 333, 2022.

\bibitem{chou2019analog}
J.~Chou, S.~Bramhavar, S.~Ghosh, and W.~Herzog, ``Analog coupled oscillator based weighted ising machine,'' \emph{Scientific reports}, vol.~9, no.~1, p. 14786, 2019.

\bibitem{delacour2023mixed}
C.~Delacour, S.~Carapezzi, G.~Boschetto, M.~Abernot, T.~Gil, N.~Azemard, and A.~Todri-Sanial, ``A mixed-signal oscillatory neural network for scalable analog computations in phase domain,'' \emph{Neuromorphic Computing and Engineering}, vol.~3, no.~3, p. 034004, 2023.

\bibitem{graber2024integrated}
M.~Graber and K.~Hofmann, ``An integrated coupled oscillator network to solve optimization problems,'' \emph{Communications Engineering}, vol.~3, no.~1, p. 116, 2024.

\bibitem{holzel2013neural}
R.~H{\"o}lzel, ``A neural network of weakly coupled nonlinear oscillators with a global, time-dependent coupling-theory and experiment,'' Ph.D. dissertation, Technische Universit{\"a}t M{\"u}nchen, 2013.

\bibitem{nikonov2020convolution}
D.~E. Nikonov, P.~Kurahashi, J.~S. Ayers, H.~Li, T.~Kamgaing, G.~C. Dogiamis, H.-J. Lee, Y.~Fan, and I.~Young, ``Convolution inference via synchronization of a coupled cmos oscillator array,'' \emph{IEEE Journal on Exploratory Solid-State Computational Devices and Circuits}, vol.~6, no.~2, pp. 170--176, 2020.

\bibitem{choi2025hardware}
\BIBentryALTinterwordspacing
W.~Choi, T.~van Bodegraven, J.~Verest, O.~Maher, D.~F. Falcone, A.~L. Porta, D.~Jubin, B.~J. Offrein, S.~Karg, V.~Bragaglia, and A.~Todri-Sanial, ``Hardware implementation of ring oscillator networks coupled by beol integrated reram for associative memory tasks,'' 2025. [Online]. Available: \url{https://arxiv.org/abs/2503.14126}
\BIBentrySTDinterwordspacing

\bibitem{torrejon2017neuromorphic}
J.~Torrejon, M.~Riou, F.~A. Araujo, S.~Tsunegi, G.~Khalsa, D.~Querlioz, P.~Bortolotti, V.~Cros, K.~Yakushiji, A.~Fukushima \emph{et~al.}, ``Neuromorphic computing with nanoscale spintronic oscillators,'' \emph{Nature}, vol. 547, no. 7664, pp. 428--431, 2017.

\bibitem{romera2018vowel}
M.~Romera, P.~Talatchian, S.~Tsunegi, F.~Abreu~Araujo, V.~Cros, P.~Bortolotti, J.~Trastoy, K.~Yakushiji, A.~Fukushima, H.~Kubota \emph{et~al.}, ``Vowel recognition with four coupled spin-torque nano-oscillators,'' \emph{Nature}, vol. 563, no. 7730, pp. 230--234, 2018.

\bibitem{buzsaki2006rhythms}
G.~Buzsaki, \emph{Rhythms of the Brain}.\hskip 1em plus 0.5em minus 0.4em\relax Oxford university press, 2006.

\bibitem{strogatz1993coupled}
S.~H. Strogatz and I.~Stewart, ``Coupled oscillators and biological synchronization,'' \emph{Scientific american}, vol. 269, no.~6, pp. 102--109, 1993.

\bibitem{lehnertz2009synchronization}
K.~Lehnertz, S.~Bialonski, M.-T. Horstmann, D.~Krug, A.~Rothkegel, M.~Staniek, and T.~Wagner, ``Synchronization phenomena in human epileptic brain networks,'' \emph{Journal of neuroscience methods}, vol. 183, no.~1, pp. 42--48, 2009.

\bibitem{malagarriga2015synchronization}
D.~Malagarriga, M.~A. Garc{\'\i}a-Vellisca, A.~E. Villa, J.~M. Buld{\'u}, J.~Garc{\'\i}a-Ojalvo, and A.~J. Pons, ``Synchronization-based computation through networks of coupled oscillators,'' \emph{Frontiers in computational neuroscience}, vol.~9, p.~97, 2015.

\bibitem{hoppensteadt1999oscillatory}
F.~C. Hoppensteadt and E.~M. Izhikevich, ``Oscillatory neurocomputers with dynamic connectivity,'' \emph{Physical Review Letters}, vol.~82, no.~14, p. 2983, 1999.

\bibitem{kuramoto1975self}
Y.~Kuramoto, ``Self-entrainment of a population of coupled non-linear oscillators,'' in \emph{International Symposium on Mathematical Problems in Theoretical Physics: January 23--29, 1975, Kyoto University, Kyoto/Japan}.\hskip 1em plus 0.5em minus 0.4em\relax Springer, 1975, pp. 420--422.

\bibitem{popescu2018simulation}
B.~Popescu, G.~Csaba, D.~Popescu, A.~H. Fallahpour, P.~Lugli, W.~Porod, and M.~Becherer, ``\BIBforeignlanguage{en}{Simulation of coupled spin torque oscillators for pattern recognition},'' \emph{\BIBforeignlanguage{en}{Journal of Applied Physics}}, vol. 124, no.~15, p. 152128, Oct. 2018.

\bibitem{rudner2024design}
T.~Rudner, W.~Porod, and G.~Csaba, ``Design of oscillatory neural networks by machine learning,'' \emph{Frontiers in Neuroscience}, vol.~18, p. 1307525, 2024.

\bibitem{follmann2014phase}
R.~Follmann, E.~Macau, E.~Rosa, and J.~Piqueira, ``Phase oscillatory network and visual pattern recognition,'' \emph{IEEE transactions on neural networks and learning systems}, vol.~26, 08 2014.

\bibitem{rumelhart1986learning}
D.~E. Rumelhart, G.~E. Hinton, and R.~J. Williams, ``Learning representations by back-propagating errors,'' \emph{nature}, vol. 323, no. 6088, pp. 533--536, 1986.

\bibitem{werbos1990backpropagation}
P.~J. Werbos, ``Backpropagation through time: what it does and how to do it,'' \emph{Proceedings of the IEEE}, vol.~78, no.~10, pp. 1550--1560, 1990.

\bibitem{kuninti2021backpropagation}
\BIBentryALTinterwordspacing
S.~Kuninti and S.~Rooban, ``Backpropagation algorithm and its hardware implementations: A review,'' \emph{Journal of Physics: Conference Series}, vol. 1804, no.~1, p. 012169, feb 2021. [Online]. Available: \url{https://dx.doi.org/10.1088/1742-6596/1804/1/012169}
\BIBentrySTDinterwordspacing

\bibitem{scellier2024energy}
B.~Scellier, M.~Ernoult, J.~Kendall, and S.~Kumar, ``Energy-based learning algorithms for analog computing: a comparative study,'' \emph{Advances in Neural Information Processing Systems}, vol.~36, 2024.

\bibitem{scellier2017equilibrium}
B.~Scellier and Y.~Bengio, ``Equilibrium propagation: Bridging the gap between energy-based models and backpropagation,'' \emph{Frontiers in computational neuroscience}, vol.~11, p.~24, 2017.

\bibitem{yi2023activity}
S.-i. Yi, J.~D. Kendall, R.~S. Williams, and S.~Kumar, ``Activity-difference training of deep neural networks using memristor crossbars,'' \emph{Nature Electronics}, vol.~6, no.~1, pp. 45--51, 2023.

\bibitem{dillavou2024machine}
S.~Dillavou, B.~D. Beyer, M.~Stern, A.~J. Liu, M.~Z. Miskin, and D.~J. Durian, ``Machine learning without a processor: Emergent learning in a nonlinear analog network,'' \emph{Proceedings of the National Academy of Sciences}, vol. 121, no.~28, p. e2319718121, 2024.

\bibitem{laydevant2024training}
J.~Laydevant, D.~Markovi{\'c}, and J.~Grollier, ``Training an ising machine with equilibrium propagation,'' \emph{Nature Communications}, vol.~15, no.~1, p. 3671, 2024.

\bibitem{nest2024towards}
T.~Nest and M.~Ernoult, ``Towards training digitally-tied analog blocks via hybrid gradient computation,'' \emph{arXiv preprint arXiv:2409.03306}, 2024.

\bibitem{wang2024training}
Q.~Wang, C.~C. Wanjura, and F.~Marquardt, ``Training coupled phase oscillators as a neuromorphic platform using equilibrium propagation,'' \emph{arXiv preprint arXiv:2402.08579}, 2024.

\bibitem{jenkins2024impact}
A.~S. Jenkins, L.~Martins, L.~C. Benetti, A.~Schulman, P.~Anacleto, M.~S. Claro, I.~Caha, F.~L. Deepak, E.~Paz, and R.~Ferreira, ``The impact of local pinning sites in magnetic tunnel junctions with non-homogeneous free layers,'' \emph{Communications Materials}, vol.~5, no.~1, p.~7, 2024.

\bibitem{tsunegi2018scaling}
S.~Tsunegi, T.~Taniguchi, R.~Lebrun, K.~Yakushiji, V.~Cros, J.~Grollier, A.~Fukushima, S.~Yuasa, and H.~Kubota, ``Scaling up electrically synchronized spin torque oscillator networks,'' \emph{Scientific reports}, vol.~8, no.~1, p. 13475, 2018.

\bibitem{rodrigues2016kuramoto}
F.~A. Rodrigues, T.~K.~D. Peron, P.~Ji, and J.~Kurths, ``The kuramoto model in complex networks,'' \emph{Physics Reports}, vol. 610, pp. 1--98, 2016.

\bibitem{slavin2009nonlinear}
A.~Slavin and V.~Tiberkevich, ``Nonlinear auto-oscillator theory of microwave generation by spin-polarized current,'' \emph{IEEE Transactions on Magnetics}, vol.~45, no.~4, pp. 1875--1918, 2009.

\bibitem{laborieux2021scaling}
A.~Laborieux, M.~Ernoult, B.~Scellier, Y.~Bengio, J.~Grollier, and D.~Querlioz, ``Scaling equilibrium propagation to deep convnets by drastically reducing its gradient estimator bias,'' \emph{Frontiers in neuroscience}, vol.~15, p. 633674, 2021.

\bibitem{ernoult2019updates}
M.~Ernoult, J.~Grollier, D.~Querlioz, Y.~Bengio, and B.~Scellier, ``Updates of equilibrium prop match gradients of backprop through time in an rnn with static input,'' \emph{Advances in neural information processing systems}, vol.~32, 2019.

\bibitem{ruder2017overview}
\BIBentryALTinterwordspacing
S.~Ruder, ``An overview of gradient descent optimization algorithms,'' 2017. [Online]. Available: \url{https://arxiv.org/abs/1609.04747}
\BIBentrySTDinterwordspacing

\bibitem{kingma2014adam}
D.~P. Kingma, ``Adam: A method for stochastic optimization,'' \emph{arXiv preprint arXiv:1412.6980}, 2014.

\bibitem{lecun1998gradient}
Y.~LeCun, L.~Bottou, Y.~Bengio, and P.~Haffner, ``Gradient-based learning applied to document recognition,'' \emph{Proceedings of the IEEE}, vol.~86, no.~11, pp. 2278--2324, 1998.

\bibitem{sakaguchi1986soluble}
H.~Sakaguchi and Y.~Kuramoto, ``A soluble active rotater model showing phase transitions via mutual entertainment,'' \emph{Progress of Theoretical Physics}, vol.~76, no.~3, pp. 576--581, 1986.

\bibitem{albertsson2021ultrafast}
D.~I. Albertsson, M.~Zahedinejad, A.~Houshang, R.~Khymyn, J.~{\AA}kerman, and A.~Rusu, ``Ultrafast ising machines using spin torque nano-oscillators,'' \emph{Applied Physics Letters}, vol. 118, no.~11, 2021.

\end{thebibliography}

\newpage
\appendix
\section*{Appendix}
\setcounter{section}{1}

\subsection{Network architecture and initialization} \label{s:network_architecture}
In this work, all the conducted simulations were done using fully-connected-layered networks with one hidden layer (also called one-hidden layer networks), composed of an input layer, one hidden layer, and an output layer. We decided to focus on this specific network architecture, mostly inspired from Deep Learning works, instead of more `physically-based' architectures such as all-to-all or nearest-neighbors couplings. In order to develop a hardware implementation of a neuromorphic system, these `physically-based' architectures would need to be deeply investigate. However, the work of this article aims to present oscillatory networks as  a candidate hardware for future neuromorphic applications. In that sense, to be able to benchmark our system with those already existing \cite{scellier2017equilibrium}\cite{ernoult2019updates}\cite{wang2024training}, we decided to use a comparable network architecture.

To train our oscillatory network, we need to encode the samples of the database we want to learn in the phase domain (see section \ref{s:data_encoding} for more details). In this work, a sample usually corresponds to an input image (grid of pixels) and a class label. To encode this kind of sample, we use oscillatory sources, i.e. oscillators for which we can tune their phase. For an input image, each source encodes one pixel. Thus, an $N\times N$ image will require $N^2$ sources to be feed to the network.
These input nodes are connected to the oscillators in the hidden layer. It corresponds to the features produced through network learning, and captures the underlying structures of the input data as well as the expressive power of the network.
Then, oscillators in the hidden layer are all connected to output oscillators that encode the prediction of the network (the label of an input image).
The whole architecture of the network is depicted in Figure \ref{f:Kuramoto_net}.
The input sources are fully-connected to hidden oscillators and continuously drive their dynamics.
Hidden and output oscillators are fully-connected in a bidirectional way and therefore, they influence each other's dynamics.

The weights between two consecutive layers $l$ and $l+1$ of a network are initialized using a uniform law $\mathcal{U}(\frac{-1}{\sqrt{N_l}}, \frac{1}{\sqrt{N_l}})$, where $N_l$ is the size of the layer $l$. When present, biases are initialized as follows: the amplitude of the bias of the layer $l$ is initialized also with $\mathcal{U}(\frac{-1}{\sqrt{N_l}})$ and its phase is initialized as a random number in the interval $[-\pi, \pi]$.

\subsection{Data encoding in the phase domain} \label{s:data_encoding}
Both input (pixels of an image) and target (label of an input image) data are encoded in the phases of sources. These sources are connected to the network and will perturb its dynamics. Through our simulations, we observed that the encoding procedure of the data may have a significant impact on the system behavior, and then, on its performance once trained with Equilibrium Propagation.
\\
We choose the following encoding procedure for the input and target data:
\begin{itemize}
    \item Input: pixels of an input image are normalized in $[-\frac{\pi}{2}, \frac{\pi}{2}]$ and phases of input sources are set to the value of the transformed pixels.
    \item Target: target labels are vectorized using one-hot encoding, and modified by the linear function $f:c\rightarrow \frac{\pi}{2}c + \frac{\pi}{2}$. Then, phases of target sources are set to the value of the transformed labels.
\end{itemize}

Both these encoding procedures have been obtained and confirmed empirically. They have been shown to give the best results and we have no theoretical proof of their optimality. However, we can give some intuitions on that.
Regarding the input encoding, we know that during its relaxation, the network tends to minimize, or maximize depending on the sign of the corresponding weight, the quantity $\cos(\phi^{\rm in}_i - \phi^{\rm hid}_h)$, i.e. make this quantity close to -1 or 1. Then, it comes that $\phi^{\rm hid}_h = \phi^{\rm in}_i - \pi$ $[2\pi]$ or $\phi^{\rm hid}_h = \phi^{\rm in}_i$ $[2\pi]$.
Hence, by choosing the previous encoding procedure, it appears that hidden phases $\phi^{\rm hid}_h$ will converge either in $[\frac{\pi}{2}, \frac{3\pi}{2}]$ or in $[-\frac{\pi}{2}, \frac{\pi}{2}]$ depending on the sign of the weight. The reunion of both intervals corresponds to the entire trigonometric circle with no overlap. If we had chosen an interval of greater length, there would have been an overlap between both intervals. If we had chosen an interval of smaller length, there would have been no overlap but the trigonometric circle would not have been covered completely. This implies that with our encoding procedure, we know the sign of the coupling weight from the phase value of the hidden oscillators. It introduces some structure inside the network that can be used during training.

Regarding the target encoding, the result was also obtained empirically. One interesting thing that we notice is that generally the network's performance are better when the phase difference in the encoding procedure is equal to $\frac{\pi}{2}$ rather than to $\pi$.

\subsection{Pseudo-code for Equilibrium Propagation} \label{s:pseudo_code_EP}

Below we give the pseudo code of the Equilibrium Propagation algorithm with three phases (prediction, positive nudge and negative nudge):
\begin{algorithm}
\caption{Equilibrium Propagation with three phases}
\begin{algorithmic}[1]
\State \textbf{Inputs:} $X_{train}, \textnormal{Model}, \beta, lr$, max\_epoch
\For{epoch $n \in \lshad1,$max\_epoch$\rshad$}
\For{each image, label $(x, l) \in X_{train}$}
\State Encode image: $x_{enc} \gets \textnormal{encode\_image}(x)$
\State Encode label: $l_{enc} \gets \textnormal{encode\_label}(l)$
\State Initialize neuron units: $\Phi \gets \textnormal{initialize}()$ \Comment{Neuron initialization}
\State \textbf{Free phase}: $\Phi^0 \gets \textnormal{Model}(x_{enc}, \Phi)$ \Comment{Compute prediction}
\State \textbf{Positive nudge phase}: $\Phi^{\beta} \gets \textnormal{Model}(x_{enc}, \Phi^0, \beta, l_{enc})$
\State \textbf{Negative nudge phase}: $\Phi^{-\beta} \gets \textnormal{Model}(x_{enc}, \Phi^0, -\beta, l_{enc})$
\State Compute gradients: $\Delta \Omega \gets$ equation (\ref{e:general_learning_rule})
\State Update weights: $\Omega \gets \Omega + lr\cdot\Delta \Omega$
\EndFor
\EndFor
\end{algorithmic}
\end{algorithm}

\subsection{Details of equations for a one hidden layer Kuramoto network}
In this section, we give details of the oscillator dynamics equations, the derivation of the network energy and the computation of the learning rules for the different learning parameters. The calculations are done for a network of Kuramoto oscillators with one hidden layer and external bias. These external signals are tunable, so they are new learning parameters in our model. All oscillators are assumed to share the same natural frequency $\omega_0$. Then, we consider the equations in the rotating frame at frequency $\omega_0$. So, we define and use the `slow phase' (phase in the rotating frame) of an oscillator $\Phi = \phi + \omega_0$ instead of the phase in the stationary frame $\phi$.

\subsubsection{Equations of dynamics}

\begin{eqnarray*}
\eqalign{
      & \frac{d\Phi^{\rm{hid}}_h}{dt} = \sum_{i=1}^I \Omega_{h,i}^0\sin(\Phi^{\rm{inp}}_i - \Phi^{\rm{hid}}_h) + \sum_{o=1}^O \Omega_{o,h}^1\sin(\Phi^{\rm{out}}_o - \Phi^{\rm{hid}}_h) \\ & \hspace{2cm} + F_h^0\sin(\Psi_{h}^0 - \Phi_h), \hspace{0.3cm} \forall h \in \lshad1,H\rshad \\
      & \frac{d\Phi^{\rm{out}}_o}{dt} = \sum_{h=1}^{H} \Omega_{o,h}^1\sin(\Phi^{\rm{hid}}_h - \Phi^{\rm{out}}_o) + F_o^1\sin(\Psi_{o}^1 - \Phi^{\rm{out}}_o), \hspace{0.3cm} \forall o \in \lshad1,O\rshad
}
\end{eqnarray*}

\noindent
where:

$(I, H, O)$: number of oscillators (or sources) in each layer, respectively input, hidden and output

$((\Phi^{\rm{inp}}_i)_{(1 \leq i \leq I)}, (\Phi^{\rm{hid}}_h)_{(1 \leq h \leq H)}, (\Phi^{\rm{out}}_o)_{(1 \leq o \leq O)})$: oscillator `slow phases' at source frequency (sources for the input), respectively in the input, hidden and output layers

$\Omega^0 = (\Omega_{h,i})_{(1 \leq h \leq H), (1 \leq i \leq I)}$: coupling matrix between input and hidden layers

$\Omega^1 = (\Omega_{o,h})_{(1 \leq o \leq O), (1 \leq h \leq H)}$: coupling matrix between hidden and output layers

$(F^0, \Psi^0) = (F^0_h, \Psi_{h}^0)_{(1 \leq h \leq H)}$: external drives (amplitude and phase) applied to oscillators in the hidden layer

$(F^1, \Psi^1) = (F^1_o, \Psi_{o}^1)_{(1 \leq o \leq O)}$: external drives (amplitude and phase) applied to oscillators in the output layer

\subsubsection{Derivation of the energy function}
From the previous dynamics, we derive the energy function of a Kuramoto network $E_{\Phi}$ and the total energy $F_{\Phi}$ influenced by `nudging' factor $\beta$.
The dynamics of the oscillators that compose the network tend to minimize the energy of the system:
\begin{eqnarray*}
\eqalign{
    \frac{\partial E_\Phi}{\partial \Phi^{\rm{inp}}_h} = -\frac{\rmd\Phi^{\rm{inp}}_h}{\rmd t}, \hspace{0.3cm} \forall h \in \lshad1,H\rshad \\
    \frac{\partial E_\Phi}{\partial \Phi^{\rm{out}}_o} = -\frac{\rmd\Phi^{\rm{out}}_o}{\rmd t}, \hspace{0.3cm} \forall o \in \lshad1,O\rshad
}
\end{eqnarray*}
\noindent

Hence,
\begin{eqnarray*}
    E_{\Phi} := -\sum_{h=1}^H \sum_{i=1}^I \Omega_{h,i}^0\cos(\Phi^{\rm{inp}}_i - \Phi^{\rm{hid}}_h) - \sum_{h=1}^H\sum_{o=1}^O \Omega_{o,h}^1\cos(\Phi^{\rm{out}}_o - \Phi^{\rm{hid}}_h) \\ - \sum_{h=1}^HF_h^0\cos(\Psi_{h}^0 - \Phi^{\rm{hid}}_h) - \sum_{o=1}^OF_o^1\cos(\Psi_{o}^1 - \Phi^{\rm{out}}_o)
\end{eqnarray*}

As described in section \ref{s:Kuramoto_EqProp}, during the nudge phase, we drive the output oscillators $\Phi^{\rm{out}}_o$ towards target sources $\Psi_{\rm{target}}$. The parameter $\beta$ controls the strength of the drive interaction. We define the total energy $F_\Phi$ of the network:
\begin{eqnarray*}
    F_{\Phi}(\Phi, \theta, \Psi_{\rm{target}}, \beta) := E_{\Phi}(\Phi, \theta, \Psi_{\rm{target}}) + \beta\mathcal{L}(\Phi, \Psi_{\rm{target}})
\end{eqnarray*}
Where

$\theta = (\Omega^0, \Omega^1, F^0, F^1, \Psi^0, \Psi^1)$: learning parameters

$\psi_{\rm{target}} = (\psi_{\rm{target},o})_{(1 \leq o \leq O)}$: source phases that encode target data

$\beta$: `nudge factor'

$\mathcal{L}$: Loss function

\subsubsection{Derivation of the learning rule}
To obtain the learning rules, we use the formula proposed in \cite{scellier2017equilibrium} that relies on the total energy of the network $F_{\phi}$ and the steady states of the network after the `free' and `nudge' phases, respectively $\Phi^{0}$ and $\Phi^{\beta}$:
\begin{eqnarray*}
    \Delta \theta_i \propto -\frac{1}{\beta} \left ( \frac{\partial F_{\Phi}}{\partial \theta_i}(\Phi^{\beta}, \theta, \Psi_{\rm{target}}, \beta) - \frac{\partial F_{\Phi}}{\partial \theta_i}(\Phi^{0}, \theta, \Psi_{\rm{target}}, 0) \right )
\end{eqnarray*}
\\
Then, we obtain the following learning rules:
\begin{eqnarray*}
\eqalign{
    & \Delta \Omega_{j,k}^0 \propto \frac{1}{\beta} \left ( \cos(\Phi^{\rm{inp},\beta}_k - \Phi^{\rm{hid},\beta}_j) - \cos(\Phi^{\rm{inp},0}_k - \Phi^{\rm{hid},0}_j) \right ) \\
    & \Delta \Omega_{j,k}^1 \propto \frac{1}{\beta} \left ( \cos(\Phi^{\rm{out},\beta}_k - \Phi^{\rm{hid},\beta}_j) - \cos(\Phi^{\rm{out},0}_k - \Phi^{\rm{hid},0}_j) \right ) \\
}
\end{eqnarray*}

\begin{eqnarray*}
\eqalign{
    & \Delta F_j^0 \propto \frac{1}{\beta} \left ( \cos(\Psi_j^0 - \Phi^{\rm{hid},\beta}_j) - \cos(\Psi_j^0 - \Phi^{\rm{hid},0}_j) \right ) \\
    & \Delta F_j^1 \propto \frac{1}{\beta} \left ( \cos(\Psi_j^1 - \Phi^{\rm{out},\beta}_j) - \cos(\Psi_j^1 - \Phi^{\rm{out},0}_j) \right )
}
\end{eqnarray*}

\begin{eqnarray*}
\eqalign{
    & \Delta \Psi_j^0 \propto \frac{F_j^0}{\beta} \left ( \sin(\Psi_j^0 - \Phi^{\rm{hid},0}_j) - \sin(\Psi_j^0 - \Phi^{\rm{hid},\beta}_j) \right ) \\
    & \Delta \Psi_j^1 \propto \frac{F_j^1}{\beta} \left ( \sin(\Psi_j^1 - \Phi^{\rm{out},0}_j) - \sin(\Psi_j^1 - \Phi^{\rm{out},\beta}_j) \right )
}
\end{eqnarray*}

\subsection{Synchronization during training}\label{s:nonsynch_coupling}
In this section, we give details and examples of the phenomenon of oscillator synchronization during training with Equilibrium Propagation, as described in Section \ref{s:nonsynch_Kuramoto}. We first explain this phenomenon with simulation results and then give an analytical demonstration using statistical tools.

\subsubsection{Empirical analysis of synchronization}\label{s:nonsynch_coupling_simulations}
To investigate oscillator synchronization during training, we simulate a network of Kuramoto oscillators and observe the behavior of one single oscillator. We set all oscillator frequencies at the same value $\omega_0$, except for one oscillator whose frequency is set to $\omega_s$ (5\% higher than $\omega_0$) so that this oscillator is not initially synchronized with the others. Then, we feed the network with one input image (taken from the Digits database) and train only one coupling weight, between the chosen oscillator and a source. Finally, we observe the evolution of the considered weight during the training and how it affects the synchronization of the oscillator.

\Figure{\label{f:synchronization_region} Evolution of the instantaneous frequency of the oscillator as a function of one coupling in a network of Kuramoto oscillators for the `free' and `nudge' phases of the Equilibrium Propagation algorithm.
\includegraphics[scale=0.7]{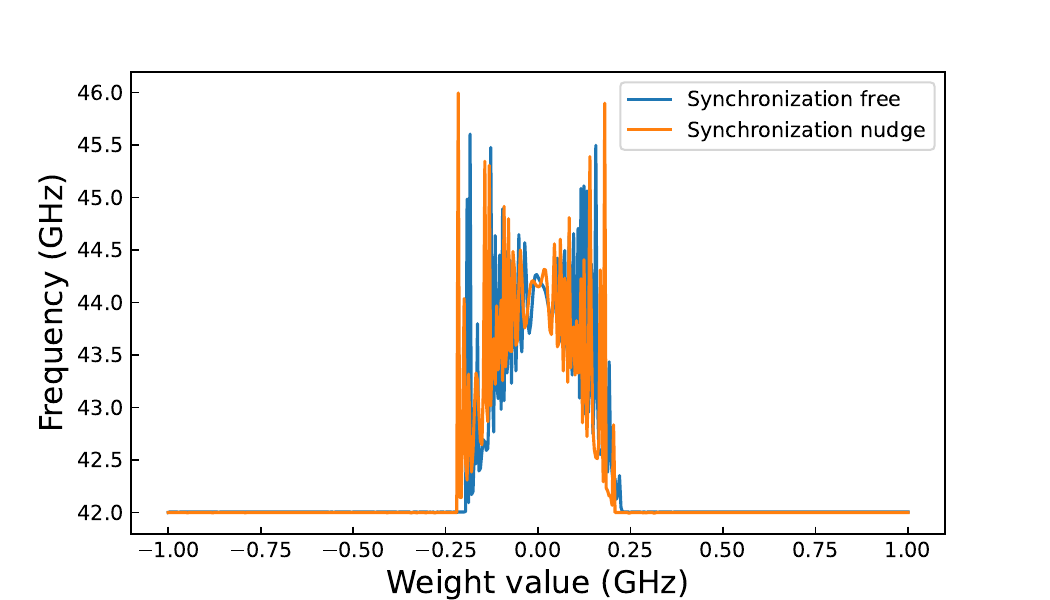}}

Figure \ref{f:synchronization_region} shows the evolution of the oscillator frequency over the coupling weight for the `free' and `nudge' steps of the Equilibrium Propagation framework. In this figure, we can see that if the weight is large enough (in absolute value), the frequency of the oscillator is equal to $\omega_0$ i.e. the oscillator synchronizes to other oscillators. Between two thresholds values, the frequency of the oscillator is not equal anymore to $\omega_0$, which means that the oscillator cannot synchronize. We distinguish the `free' and `nudge' steps of the Equilibrium Propagation framework because the network is not driven by the same elements in each case, hence it slightly changes its synchronization region.

\Figure{\label{f:synchronization_during_training} Evolution of the coupling weight (left) and weight update computed by the learning rule \ref{e:synch_gradients} (right) as a function of the number of updates for a single weight in a one hidden layer network with 50 hidden oscillators. Using different colors, we recorded the different synchronization states of the oscillator. The color blue means that the considered oscillator is not synchronized at the frequency $\omega_0$. The color red means synchronization only during the `free' phase. The color green means synchronization only during the `nudge' phase. Finally, the color orange means synchronization during both `free' and `nudge' phases. We used $\beta=1$ and $lr=0.12$.
\includegraphics[scale=0.5]{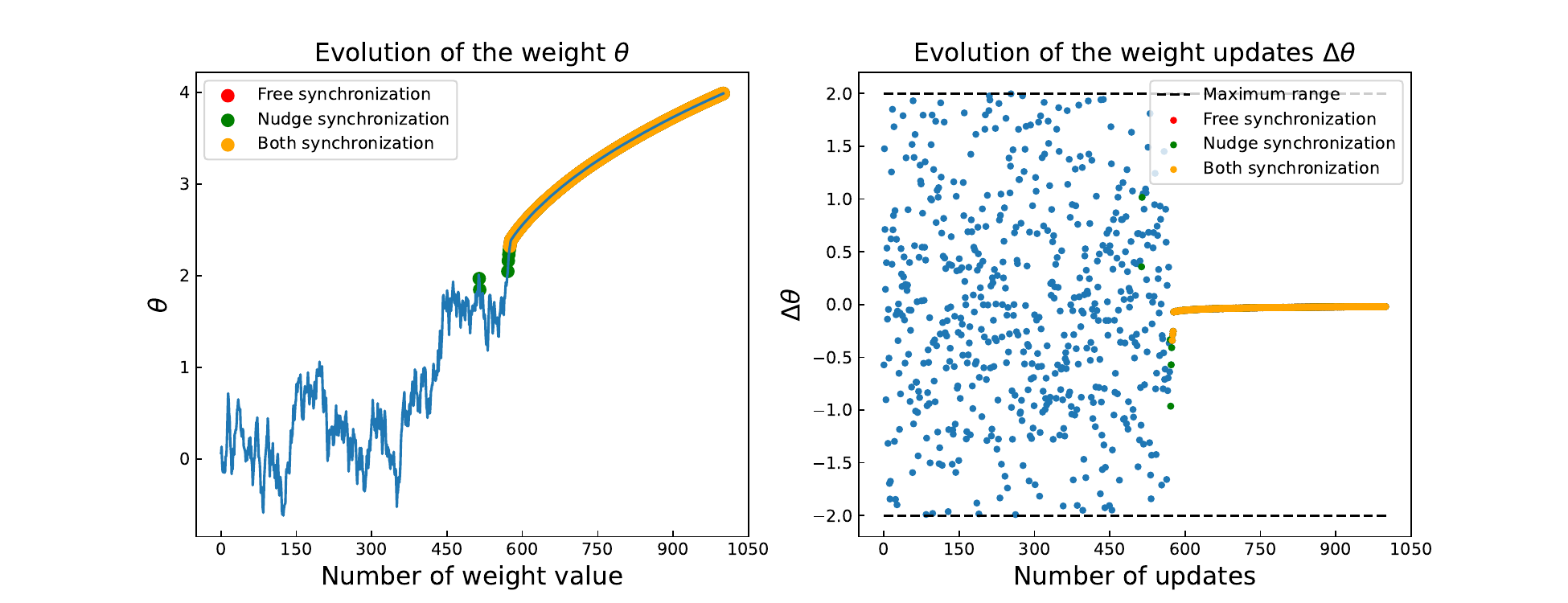}}

\Figure{\label{f:histogram_weight_update} Histogram of the amplitude of the weight updates during training for one single oscillator in a one hidden layer network with 50 hidden oscillators. We split the weight updates using two different colors, depending on the synchronization state of the considered oscillator. The color blue corresponds to weight updates before the oscillator reaches a synchronization state (synchronization at $\omega_0$ for both 'free' and 'nudge' phases), whereas the orange color corresponds to the weight updates after the oscillator reaches the synchronization state. We used $\beta=1$ and $lr=0.12$.
\includegraphics[scale=0.6]{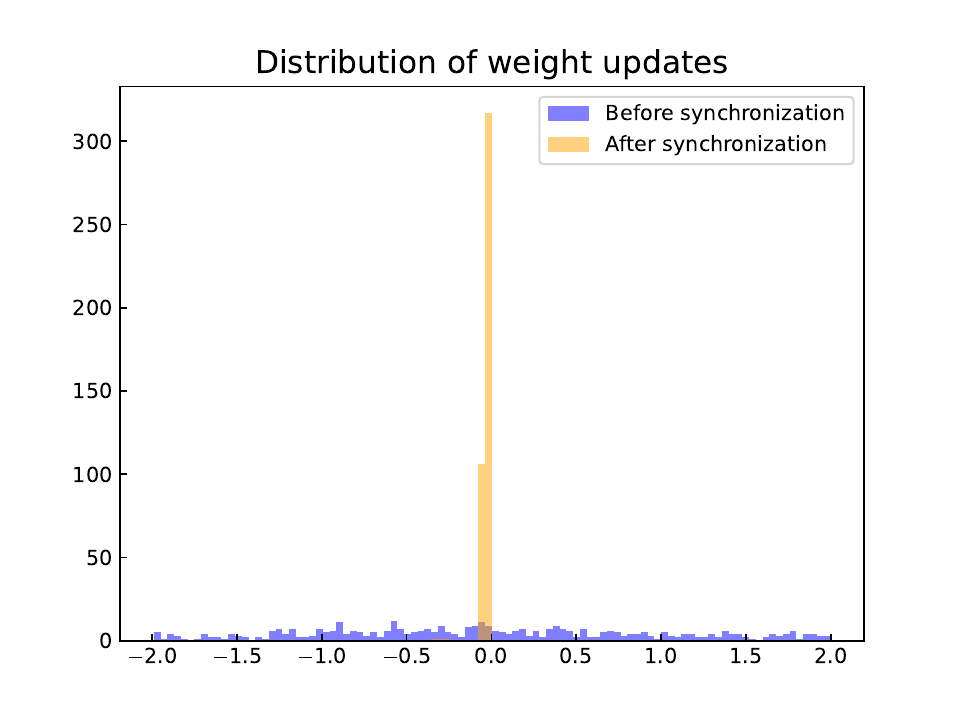}}

Figure \ref{f:synchronization_during_training} shows on the left the evolution of the weight during training, and on the right the value of each weight update. In both of these plots, we highlight the synchronization state of the oscillator for each step of the Equilibrium Propagation framework (`free' and `nudge'), and for both. Dashed black lines on the right show the maximum range of update that can be reached.
On the left, we observe that the value of the weight is evolving in a random way until it reaches a value that enables the synchronization of the oscillator. Once the oscillator reaches this synchronization state, the value of the weight stabilizes and starts a slow convergence.
On the right, we can see that first the weight updates seem to follow a random walk on the interval $[-\frac{2}{\beta}, \frac{2}{\beta}]$. Then, once the oscillator reaches a synchronized state, we observe that weight updates change radically their behavior and start a slow convergence towards 0. This case corresponds to the learning of the optimal weight regarding the initial task. Another interesting point to notice is that the amplitude of the weight update also changes radically between the non-synchronized and synchronized states. In fact, as shown in the histogram \ref{f:histogram_weight_update}, before synchronization, the update follows a random path and covers the entire range of possible values. Then, when the oscillator reaches synchronization, the amplitude of the weight update decreases and starts converging towards 0.
Hence, on the one hand, if two oscillators are not synchronized, the coupling between them will evolve randomly with a large amplitude, so it will relatively quickly reach the synchronization region. On the other hand, once the coupling is in the synchronization region, its update amplitude will be smaller and deterministic, and so it has more chance to remain in the synchronization region.
This fact explains why oscillators globally synchronize during training with Equilibrium Propagation.

\subsubsection{Statistical analysis of synchronization}\label{s:nonsynch_coupling_analysis}

\Figure{\label{f:CDF_dist} Cumulative distribution function (CDF) of the Arcsine law (dashed red line) and both empirical distribution functions (EDF) of the random variables $X^0_{j,k} = \cos(\phi_j^0 - \phi_k^0)$ (red line) and $X^\beta_{j,k} = \cos(\phi_j^\beta - \phi_k^\beta)$ (green line) for a one hidden layer network with 50 hidden oscillators. We used $\beta=0.5$ and $lr=0.001$ and 1000 observations of $X^0_{j,k}$ and $X^\beta_{j,k}$ 
\includegraphics[scale=0.7]{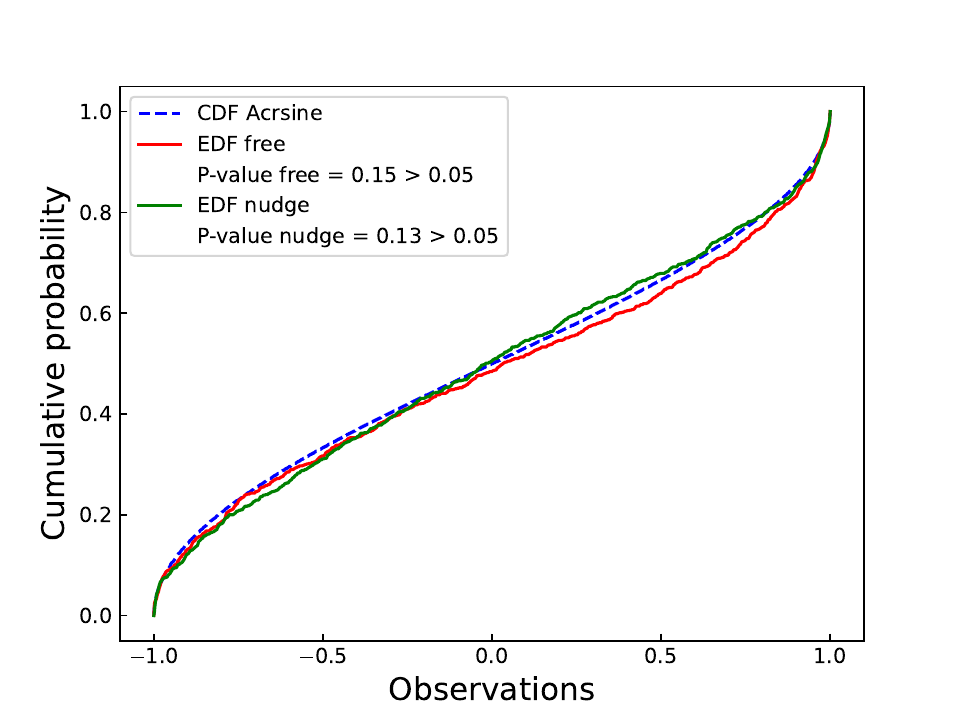}}

Using some statistical tools, we found that both cosine terms $\cos(\phi_k - \phi_j)$ in the learning rule follow an Arcsine distribution of probability (as $\phi_k - \phi_j$ follows a uniform distribution in the interval $[0,2\pi]$). We confirm this theoretical result with simulation using a Kolmogorov–Smirnov (KS) test. To do that, we consider that both cosine terms are random variables. We denote these variables $X^0_{j,k} = \cos(\phi_j^0 - \phi_k^0)$ and $X^\beta_{j,k} = \cos(\phi_j^\beta - \phi_k^\beta)$. The null hypothesis of the test is that both $X^0_{j,k}$ and $X^\beta_{j,k}$ follow a Arcsine law.
After generating 1000 samples of $X^0_{j,k}$ and $X^\beta_{j,k}$, the statistical test finally indicates that for both $X^0_{j,k}$ and $X^\beta_{j,k}$, it is impossible to reject the null hypothesis at level $\alpha=0.05$ (common level for this kind of test).

Figure \ref{f:CDF_dist} shows the cumulative distribution function (CDF) of the Arcsine law (dashed red line) and both empirical distribution functions (EDF) of the random variables $X^0_{j,k}$ (red line) and $X^\beta_{j,k}$ (green line). To obtain this plot, we simulate a one hidden layer network with 50 hidden oscillators and train only one synapse for 1000 iterations. We added the obtained P-values of the KS test to the legend of the plot. It supports the fact that both $X^0_{j,k}$ and $X^\beta_{j,k}$ follow an Arcsine law of probability.
Therefore, the coupling update $\Delta \Omega_{j,k}$ follows a difference of two Arcsine distribution, renormalized by the parameter $\beta$, i.e. $\Delta \Omega_{j,k} = \frac{1}{\beta}(X^0_{j,k} - X^\beta_{j,k})$ with $X^0_{j,k}, X^\beta_{j,k} \sim Arcsin(-1, 1)$.

From this point, we deduce theoretical estimates of the mean and variance of the distribution of $\Delta \Omega_{j,k}$:
\begin{eqnarray*}
\eqalign{
    & \mathrm{E}\left[\Delta \Omega_{j,k}\right] = 0 \\
    & \mathrm{V}\left[\Delta\Omega_{j,k}\right] = \frac{1}{\beta^2}\\
}
\end{eqnarray*}

Thus, the variance of the update diverges as we decrease the `nudging factor' $\beta$, leading to extreme values of $\Delta \Omega_{j,k}$. At the same time, the theory of Equilibrium Propagation, described in \cite{scellier2017equilibrium} assures that, when $\beta$ goes to 0, the weight update $\Delta \Omega_{j,k}$ converges towards the gradient of the loss function $\frac{\rmd \mathcal{L}}{\rmd \Omega_{j,k}}$, which is a finite value (since our loss function is differentiable). Therefore, by using sufficiently `small' values of $\beta$ (which in practice has never appeared to be a issue), we obtain a situation where the majority of the random non-synchronized weight updates are larger in amplitude than the deterministic synchronized weight updates.
This explains why the amplitude of the updates is greater than in the synchronized case, as shown empirically in Figure \ref{f:histogram_weight_update}.

\subsection{Equations of dynamics for a network of nonlinear oscillators}
In this section we give more details of the dynamics equations for a network of $K$ oscillators following the nonlinear model proposed in \cite{slavin2009nonlinear}, and on the impact of several physical parameters.

\subsubsection{The nonlinear oscillator model in a large array of oscillators}
We consider a system of $K$ coupled oscillators defined by the complex variable $c_j$, with power $p_j$ and phase $\phi_j$ influenced by an external drive:
\begin{eqnarray} \label{eq1}
    \frac{\rmd c_j}{\rmd t} + i\omega_j(p_j)c_j + \left( \Gamma_+(p_j) - \Gamma_-(p_j) \right)c_j = \sum_{k=0}^K \Omega_{k,j}e^{i\beta_{k,j}}c_k + F_je^{i\psi_j(t)}
\end{eqnarray}
where:

$c_j = \sqrt{p_j}e^{i\phi_j}$: state of oscillator $j$

$\omega_j(p) = \omega_{0,j} + Np$: frequency of oscillator $j$ 

$\Gamma_+(p) = \Gamma_G(1 + Qp)$: positive damping 

$\Gamma_-(p) = \sigma I(1 - p)$: negative damping 

$\Omega_{k,j}e^{i\beta_{k,j}}$: coupling term between oscillators $k$ and $j$

$F_je^{i\psi_j(t)}$: external drive (bias) applied to oscillator $j$

\subsubsection{Deriving power and phase dynamic equations} \label{s:derivation_Slavin}
From the previous equation in the complex domain, we can derive the coupled system of equations for the power and phase of each oscillator:

\begin{eqnarray*}
\frac{\rmd c_j(t)}{\rmd t} & = \frac{\rmd}{\rmd t} \sqrt{p_j(t)}\rme^{i\phi_j(t)}\\
 & = \frac{1}{2\sqrt{p_j}}\frac{\rmd p_j}{\rmd t}e^{i\phi_j} + i\sqrt{p_j}\frac{\rmd\phi_j}{\rmd t}\rme^{i\phi_j}
\end{eqnarray*}
\\
Then dividing by $e^{i\phi_j}$ in both terms,
\begin{eqnarray*} \label{eq3}
\frac{1}{2\sqrt{p_j}}\frac{\rmd p_j}{\rmd t} + i\sqrt{p_j}\frac{\rmd\phi_j}{\rmd t} + i\omega_j(p_j)\sqrt{p_j} + \left( \Gamma_+(p_j) - \Gamma_-(p_j) \right)\sqrt{p_j} \nonumber\\ \hspace{2cm} = \sum_{k=1}^K \Omega_{k,j}\sqrt{p_k}\rme^{i(\phi_k - \phi_j + \beta_{k,j})} + F_j\rme^{i(\psi_j - \phi_j)}
\end{eqnarray*}
\\
By taking real and imaginary parts and isolating the derivative terms we obtain,
\begin{eqnarray*}
\eqalign{
    & \frac{\rmd p_j}{\rmd t} + 2p_j\left( \Gamma_+(p_j) - \Gamma_-(p_j) \right) = 2\sum_{k=1}^K \Omega_{k,j}\sqrt{p_j p_k}\cos(\phi_k - \phi_j + \beta_{k,j}) \\ & \hspace{2cm} + 2F_j\sqrt{p_j}\cos(\psi_j - \phi_j)\\
    & \frac{\rmd\phi_j}{\rmd t} + \omega_j(p_j) = \sum_{k=1}^K \Omega_{k,j}\sqrt{\frac{p_k}{p_j}}\sin(\phi_k - \phi_j + \beta_{k,j}) + \frac{F_j}{\sqrt{p_j}}\sin(\psi_j - \phi_j)
}
\end{eqnarray*}

In the phase equation, we recognize several terms that appear in the Kuramoto dynamics (see equation \ref{e:nonsynch_Kuramoto}) such as the pairwise couplings $\Omega_{i,j}$ and the frequency term $\omega_j$. However, in this case, both of these terms also depend on the power $p_j$ of the considered oscillator.
The coupling term is weighted by the ratio of amplitudes of the coupled oscillators.
A term proportional to the oscillator power is added to the frequency term. The proportionality constant $N$ is the nonlinear frequency shift, a physical parameter that depends on the type of oscillator we are considering. We also note the presence of coupling phases $\beta_{k,j}$ within the coupling dynamics. This term influences the synchronization of the system and is closely related to the nonlinear frequency shift $N$ in the frequency term $\omega_j(p_j)$ (you can find more information on the links between the physical parameters $N$ and $\beta_{k,j}$ and their impacts in \ref{s:nonlinear_frequency_shift}).

The power equation is defined by similar physical influences as the pairwise coupling. Moreover, it also depends on a damping term which is the difference between a positive damping term (natural energy dissipation) and a negative damping term (energy supply from the external energy source, i.e. input current). From this term we derive the minimum amount of injected current to make our system operate at a positive power as described in \cite{slavin2009nonlinear}.

To use the framework provided by Equilibrium Propagation with the previous system of nonlinear oscillators, we introduce an energy function that derives from the previous dynamics.
As for the Kuramoto model, the oscillators are coupled in phase, so we consider an energy $E_\phi$ that derives from the phase dynamics of equations \ref{e:Slavin_dynamics}:

\begin{eqnarray*}
\fl E_\phi(\phi, p, \theta, \psi_{\rm{target}}, \beta) = \sum_{j=1}^K \omega_j(p_j)\phi_j - \sum_{j=1}^K \sum_{k=1}^K \Omega_{k,j}\sqrt{\frac{p_k}{p_j}} \cos(\phi_k - \phi_j + \beta_{k,j}) \nonumber - 
\sum_{j=1}^K\frac{F_j}{\sqrt{p_j}}\cos(\psi_j - \phi_j)
\end{eqnarray*}

To obtain this energy function, we did not take into account the fact that oscillator powers are also evolving during the stabilization of the system. However, we observe in our simulations that oscillator powers converge towards their steady states in a shorter timescale than the phase of the oscillators (this is particularly true when we increase the oscillator input power). Then, this phenomenon makes this approximation relevant.

We can derive the total energy $F_\phi$ of the system:
\begin{eqnarray*}
    F_{\phi}(\phi, p, \theta, \psi_{\rm{target}}, \beta) := E_{\phi}(\phi, p, \theta, \psi_{\rm{target}}) + \beta\mathcal{L}(\phi, \psi_{\rm{target}})
\end{eqnarray*}
With,

$\phi = (\phi_j)_{(1 \leq j \leq K)}$: oscillator phases

$p = (p_j)_{(1 \leq j \leq K)}$: oscillator powers

$\theta = (\Omega, F, \psi)$: learning parameters

$\psi_{\rm{target}} = (\psi_{\rm{target},o})_{(1 \leq o \leq O)}$: source phases that encode the target

$\beta$: nudge factor

We decided to use a loss function related to the one used in the Kuramoto model, but we also consider the influence of the oscillator power in the dynamics. Then, the loss function we use is weighted by the ratio between the amplitude of the target source $\sqrt{p_{\rm{target}, o}}$ and the amplitude of the corresponding output oscillator $\sqrt{p_o}$:
\begin{eqnarray*}
    \mathcal{L} := \sum_{o=1}^O \sqrt{\frac{p_{\rm{target}, o}}{p_o}} \cos(\psi_{\rm{target},o} - \phi_o)
\end{eqnarray*}

If we note $(p^0, \phi^0)$ the states of the oscillators after the `free phase' and $(p^\beta, \phi^\beta)$ the state of the oscillators after the `nudge phase', we can obtain the learning rule for each learning parameter $(\Omega, F, \psi)$ that will be used during training:
\begin{eqnarray*}
\eqalign{
      & \Delta \Omega_{k,j} \propto \frac{1}{\beta} \left( \sqrt{\frac{p_k^{\beta}}{p_j^{\beta}}}\cos(\phi_k^{\beta} - \phi_j^{\beta} + \beta_{k,j}) - \sqrt{\frac{p_k^0}{p_j^0}}\cos(\phi_k^0 - \phi_j^0 + \beta_{k,j}) \right) \\
      & \Delta F_j \propto \frac{1}{\beta} \left( \frac{1}{\sqrt{p_j^{\beta}}}\cos(\psi_{j} - \phi_j^{\beta}) - \frac{1}{\sqrt{p_j^0}}\cos(\psi_{j} - \phi_j^0) \right) \\
      & \Delta \psi_{j} \propto \frac{F_j}{\beta} \left( \frac{1}{\sqrt{p_j^0}}\sin(\psi_{j} - \phi_j^0) - \frac{1}{\sqrt{p_j^{\beta}}}\sin(\psi_{j} - \phi_j^{\beta}) \right)
}
\end{eqnarray*}

\subsubsection{Equations for a one layer network} Now we give the full equations for a one hidden layer network:

\begin{eqnarray*}
\eqalign{
      & \frac{\rmd p^{\rm{hid}}_h}{\rmd t} + 2p^{\rm{hid}}_h\left( \Gamma_+(p^{\rm{hid}}_h) - \Gamma_-(p^{\rm{hid}}_h) \right) = 2\sum_{i=1}^I \Omega^0_{h,i}\sqrt{p^{\rm{hid}}_h p^{\rm{inp}}_i}\cos(\phi^{\rm{inp}}_i - \phi^{\rm{hid}}_h + \beta_{i,h}) \\ & \hspace{2cm} + 2\sum_{o=1}^O \Omega^1_{o,h}\sqrt{p^{\rm{hid}}_h p^{\rm{out}}_o}\cos(\phi^{\rm{out}}_o - \phi^{\rm{hid}}_h + \beta_{o,h}) \\ & \hspace{2cm} + 2F^0_h\sqrt{p^{\rm{hid}}_h}\cos(\psi^0_{h} - \phi^{\rm{hid}}_h), \hspace{0.3cm} \forall h \in \lshad1,H\rshad\\
      & \frac{\rmd\phi^{\rm{hid}}_h}{\rmd t} + \omega_h(p^{\rm{hid}}_h) = \sum_{i=1}^I \Omega^0_{h,i}\sqrt{\frac{p^{\rm{inp}}_i}{p^{\rm{hid}}_h}}\sin(\phi^{\rm{inp}}_i - \phi^{\rm{hid}}_h + \beta_{i,h}) \\ & \hspace{2cm} + \sum_{o=1}^O \Omega^1_{o,h}\sqrt{\frac{p^{\rm{out}}_o}{p^{\rm{hid}}_h}}\sin(\phi^{\rm{out}}_o - \phi^{\rm{hid}}_h + \beta_{o,h}) \\ & \hspace{2cm} + \frac{F^0_h}{\sqrt{p^{\rm{hid}}_h}}\sin(\psi^0_{h} - \phi^{\rm{hid}}_h), \hspace{0.3cm} \forall h \in \lshad1,H\rshad
      \\
      & \frac{\rmd p^{\rm{out}}_o}{\rmd t} + 2p^{\rm{out}}_o\left( \Gamma_+(p^{\rm{out}}_o) - \Gamma_-(p^{\rm{out}}_o) \right) = 2\sum_{h=1}^H \Omega^1_{o,h}\sqrt{p^{\rm{out}}_h p^{\rm{hid}}_i}\cos(\phi^{\rm{hid}}_h - \phi^{\rm{out}}_o + \beta_{h,o}) \\ & \hspace{2cm} + 2F^1_o\sqrt{p^{\rm{out}}_o}\cos(\psi^1_{o} - \phi^{\rm{out}}_o), \hspace{0.3cm} \forall o \in \lshad1,O\rshad \\
      & \frac{\rmd\phi^{\rm{out}}_o}{\rmd t} + \omega_o(p^{\rm{out}}_o) = \sum_{h=1}^H \Omega^1_{o,h}\sqrt{\frac{p^{\rm{hid}}_h}{p^{\rm{out}}_o}}\sin(\phi^{\rm{hid}}_h - \phi^{\rm{out}}_o + \beta_{h,o}) \\ & \hspace{2cm} + \frac{F^1_o}{\sqrt{p^{\rm{out}}_o}}\sin(\psi^1_{o} - \phi^{\rm{out}}_o), \hspace{0.3cm} \forall o \in \lshad1,O\rshad
}
\end{eqnarray*}

\noindent
where:

$(I, H, O)$: number of oscillators (or sources) in each layer, respectively input, hidden and output layers

$((p^{\rm{inp}}_i)_{(1 \leq i \leq I)}, (p^{\rm{hid}}_h)_{(1 \leq h \leq H)}, (p^{\rm{out}}_o)_{(1 \leq o \leq O)})$: oscillator power (sources for the input), respectively in the input, hidden and output layers

$((\phi^{\rm{inp}}_i)_{(1 \leq i \leq I)}, (\phi^{\rm{hid}}_h)_{(1 \leq h \leq H)}, (\phi^{\rm{out}}_o)_{(1 \leq o \leq O)})$: oscillator phases in the stationary frame (sources for the input), respectively in the input, hidden and output layers

$\Omega^0 = (\Omega_{h,i})_{(1 \leq h \leq H), (1 \leq i \leq I)}$: coupling matrix between input and hidden layers

$\Omega^1 = (\Omega_{o,h})_{(1 \leq o \leq O), (1 \leq h \leq H)}$: coupling matrix between hidden and output layers

$(F^0, \psi^0) = (F^0_h, \psi_{h}^0)_{(1 \leq h \leq H)}$: external drives (amplitude and phase) applied to oscillators in the hidden layer

$(F^1, \psi_{\rm{ext}}^1) = (F^1_o, \psi_{\rm{ext},o}^1)_{(1 \leq o \leq O)}$: external drives (amplitude and phase) applied to oscillators in the output layer

\subsubsection{Nonlinear frequency shift $N$ and coupling phase $\beta_{k,j}$} \label{s:nonlinear_frequency_shift}
A very particular aspect of the nonlinear oscillator model of oscillators is the dependency between an oscillator frequency and its power. In fact, a term proportional to the oscillator power is added to the natural frequency of the oscillator. The standard frequency term $\omega_0$ in \ref{e:nonsynch_Kuramoto} becomes $\omega_0 + Np$ with $p$ the oscillator power and $N$ the nonlinear frequency shift, a intrinsic parameter of the considered oscillator.
In the theoretical framework exposed in \cite{slavin2009nonlinear}, the authors showed in case of two coupled oscillators that this parameter has a direct impact on the size of the synchronization region of the system.

To achieve a maximum size of the synchronization bandwidth, authors in \cite{slavin2009nonlinear} also explained that we need to choose carefully the coupling phases $(\beta_{k,j})_{1 \leq k,j \leq K}$ that appears in the coupling dynamics. Physically, the value of these parameters depends on the coupling mechanism or the time delay between the oscillators. For two almost identical nonlinear oscillators, they showed that a careful choice of coupling phases can increase the size of the synchronization bandwidth by compensating for the frequency shift of the oscillators. In that specific case, we would need to choose the coupling phases as:
\begin{eqnarray} \label{e:coupling_phases}
    \beta_{k,j} = \beta = \arctan(\nu) = \arctan\left(\frac{N}{\Gamma_G(Q + I_{ratio})}\right)
\end{eqnarray}
To our knowledge, there is no analytic relation between the coupling phases and the size of the locking bandwidth in the case of an arbitrary number of such oscillators with random pairwise couplings $\Omega_{k,j}$. For the Kuramoto model, it has been shown in \cite{sakaguchi1986soluble} that in some particular conditions, it is possible to obtain an analytic expression for the number of synchronized oscillators as a function of the coupling phases. But these results do not clearly show that the synchronization is maximum for a particular value of the coupling phases.
If we use the value of equation \ref{e:coupling_phases} in the case of a large network of oscillators, it appears that the nonlinear model can be reduced to the fully-synchronized Kuramoto model. With this result, we can provide a fair comparison between both models as showed in Table \ref{t:comparison_Ising_Kuramoto}.

As mentioned in \cite{slavin2009nonlinear}, by introducing the effective oscillator phases, $\tilde{\phi}_j$ that depends on the nonlinear frequency shift and the oscillator power, the nonlinear oscillator model can be reduced to the following system of equations:
\begin{eqnarray} \label{e:reduction_spintorque}
    \frac{\rmd \tilde{\phi}_j}{\rmd t} + \omega_{\rm{g},j} = \sum_{k=1}^K \tilde{\Omega}_{k,j} \sin(\tilde{\phi}_k - \tilde{\phi}_j + \tilde{\beta}_{k,j}), \hspace{0.3cm} \forall j \in \lshad1,K\rshad
\end{eqnarray}
with $\omega_{\rm{g},j}$ the free-running frequency of the $j$th oscillator, $\tilde{\Omega}_{k,j}$ and $\tilde{\beta}_{k,j}$ the renormalized couplings frequencies and phases influenced by the nonlinear frequency shift.

Equation \ref{e:reduction_spintorque} corresponds to the Kuramoto model with the coupling phases $\tilde{\beta}_{k,j} = \beta_{k,j} - \arctan(N/(\Gamma_G(Q + I_{ratio})))$. Then, using equation \ref{e:coupling_phases}, we obtain a renormalized coupling phases equal to 0. Hence equation \ref{e:reduction_spintorque} corresponds to the Kuramoto model described in section \ref{s:nonsynch_Kuramoto}.

\subsection{Simulations details} \label{s:simulation_details}
In this section, we provide several information related to the simulations carried out in this work.

\subsubsection{Physical framework}
All the values of the physical quantities used in this work (frequency, couplings, damping constant) are inspired from the field of spintronics \cite{albertsson2021ultrafast}.
In particular, we consider sources that emit at a frequency $\omega_0 = 4.2$ GHz and the couplings between oscillators are of the order of a hundred MHz.

\subsubsection{Simulation framework}
All the simulations have been carried out using the Pytorch framework of Python. The largest simulations of both models (Kuramoto and nonlinear models) have been run on two to four GPUs V100 operating in parallel. To solve the dynamics equations (which consumed most of the resources and time), we used second-order Runge-Kutta numerical methods.

\subsubsection{Simulation issues} \label{s:simulation_issues}

The results on MNIST with a network of oscillators following the nonlinear universal model were quite challenging to obtain since both the physical system and the Equilibrium Propagation algorithm require high computational power.
We also encountered several simulation issues. In fact, the simulated model is composed of two coupled equations (power and phase) with the power variable that needs to remain positive. In case of a large number of oscillators, we observed that several of them may collapse instead of reaching their positive final value. This is due to the coupling interactions between oscillators which can greatly affect their behaviors. This has often led to the crash of the simulations. To prevent this phenomenon, we increase the amount of the current injected into the oscillators. For a high quantity of injected current, we observe a more stable power dynamics which directly implies an increase of the final accuracy.

Simulations with the Kuramoto model were easier to handle. In that case, oscillators are only described with one variable (their phase). Then for small networks (a few dozen of coupled oscillators), it was possible to carry out training on a CPU. However for larger network we were forced to increase precision when solving the equations of dynamics, which required the use of GPUs.

\subsubsection{Database}
All conducted simulations were carried out on the MNIST database \cite{lecun1998gradient} and its rescaled version provided by the scikit-learn package of Python. The original database is divided into training and test sets, that are respectively composed of 60000 and 10000 $28\times28$ images of greyscale handwritten digits. The rescaled database is composed of one set of 1797 $8\times8$ images (\url{https://scikit-learn.org/stable/modules/generated/sklearn.datasets.load_digits.html}).

\subsubsection{Choice of Hyperparameters}\label{s:hyperparameters}
\begin{table}[h]
\caption{\label{t:hyperparameters}Hyperparameters used for training. `Hidden' indicates the number of hidden units, $N_{\rm free}$ and $N_{\rm nudge}$ correspond respectively to the number of steps to solve the dynamics of the system for the `free' and `nudge' phases, with the resolution step $\epsilon$, and $\beta$ denotes the value of the nudge factor.}
\begin{indented}
\item[]\begin{tabular}{@{}llllllll}
\br
Model&Hidden&$N_{\rm free}$&$N_{\rm nudge}$&$\epsilon$&$\beta$&Epochs& Learning rates\\
\mr
Kuramoto &500&1500&1000&0.01&0.1&100&0.001-0.0005\\
Non-linear &500&1500&1000&0.01&0.1&40&0.0005-0.00025\\
\br
\end{tabular}
\end{indented}
\end{table}
The description of the hyperparameters used in the conducted simulations is given in Table \ref{t:hyperparameters}. To choose the hyperparameters of Equilibrium Propagation ($N_{\rm free}$, $N_{\rm nudge}$, $\epsilon$ and $\beta$), we tried to understand which combination of these parameters give a good quality gradient approximation. In fact in \cite{ernoult2019updates}, the authors demonstrate theoretically and experimentally that gradients computed by Equilibrium Propagation tend to approximate those calculated by the Back-Propagation Through Time algorithm (BPTT) for vanishing values of the `nudge factor' $\beta$. Hence, we compared the gradients produced by Equilibrium Propagation and those by BPTT for different hyperparameters.

\Figure{\label{f:cosine_sim_Kuramoto} Average cosine similarity between gradients produced by EP (P-EP and C-EP denote respectively Positive-EP and Centered-EP, denomination used in \cite{scellier2024energy}) and BPTT as a function of the nudge factor $\beta$, for a Kuramoto model with $N_{\rm free} = 2000$, $N_{\rm nudge} = 500$, $\epsilon = 0.2$. The average is done over 200 random images of MNIST database.
\includegraphics[scale=0.6]{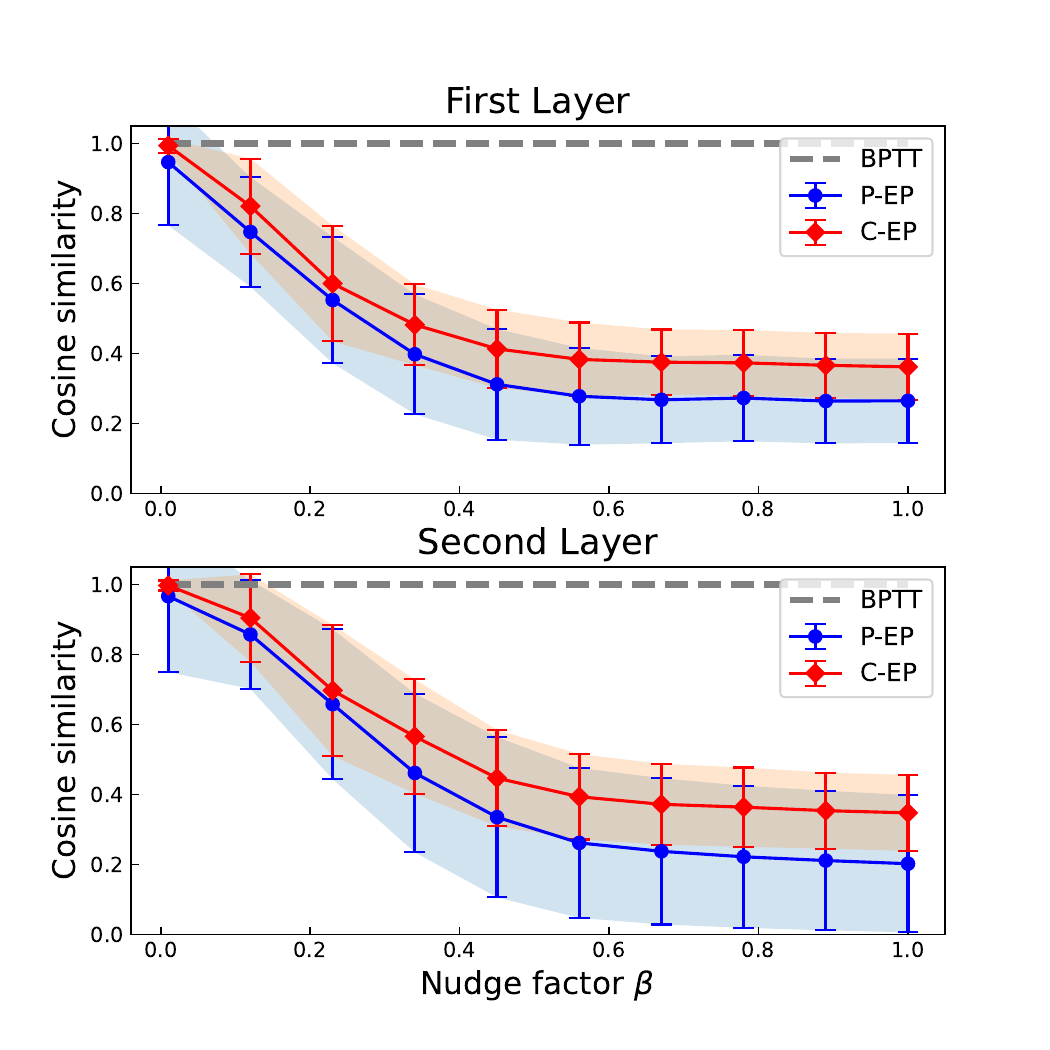}}

Figure \ref{f:cosine_sim_Kuramoto} shows the evolution of the cosine similarity between the gradients computed by EP (P-EP and C-EP denote respectively Positive-EP and Centered-EP, denomination used in \cite{scellier2024energy}) and those computed by BPTT for different values of the `nudge factor' $\beta$. As expected, the metric is maximal in both layer for very small values of $\beta$. As we increase its value, the cosine similarity decreases, which means that the direction of the computed gradients deviates from the one computed by BPTT. This result has been confirmed through simulations, i.e. "large" values of $\beta$ usually lead to poor performance of the model. As proposed in Figure \ref{f:cosine_sim_Kuramoto}, we decided to choose $\beta = 0.1$, a relatively small value of the nudge factor, to provide good gradient approximations. We observe through our simulations that smaller values of $\beta$ does not really increase the performances of the models.

For the other parameters ($N_{\rm free}$, $N_{\rm nudge}$ and $\epsilon$), we were facing a trade-off between the quality of approximation and the computation time. To obtain better approximations, we should increase $N_{\rm free}$ and $N_{\rm nudge}$, and decrease $\epsilon$. However, this also increases computation times because the relaxation of the oscillator is slower. In this situation, we observe that the quality of the gradients increases after a few iterations. This can be explained by the fact that initially the weights of the network are random, so the order of magnitude of the computed gradients is larger than after a few iterations, once the weights have been updated in the right direction. It implies that the precision used to solve the system dynamics can be a little bit reduced. After several simulations, we choose the parameters given in Table \ref{t:hyperparameters}.
Regarding the number of epochs, we have been constraint by the computation time required by the nonlinear oscillator model. In fact, this model involves two coupled variables (power and phase of an oscillator), so its computation time is greater than the Kuramoto model (that involves only one variable). We were able to train this model with sufficient precision only for 40 epochs (and 100 epochs for Kuramoto).
Finally, the choice of the learning rates was relatively straightforward for both models by observing the error and loss curves.

\subsubsection{Optimization}
Model optimization was performed using Adam optimizer \cite{kingma2014adam} with batches of size 64 and a learning rate scheduler that multiply the current learning rates by 0.98 at the end of each training epoch. Moreover we set the learning rate of the bias phase to 100 times the one of the bias amplitude.

\subsection{Impact of the bias on the learning} \label{s:bias_impact}
We tried to quantify the impact of the external drives applied to oscillators (called bias in Machine Learning). In the perspective of a hardware implementation, this question can be relevant because it changes the total number of parameters in the system. To do that, we compare the performance of two models of fully-synchronized Kuramoto oscillators, one including external drives and the other not. We compare these models on Digits and MNIST databases. For Digits database ($8\times8$ pixels per image), we use a one-hidden layer network with 50 hidden neurons. For MNIST ($28\times28$ pixels per image), we use 500 hidden neurons.

\Figure{\label{f:Bias_impact_Digits_MNIST} Average loss curves in logarithmic scale (mean over 3 runs) for a Kuramoto network trained on Digits database (left) and MNIST database (right) with and without external drives (bias).
\includegraphics[scale=0.5]{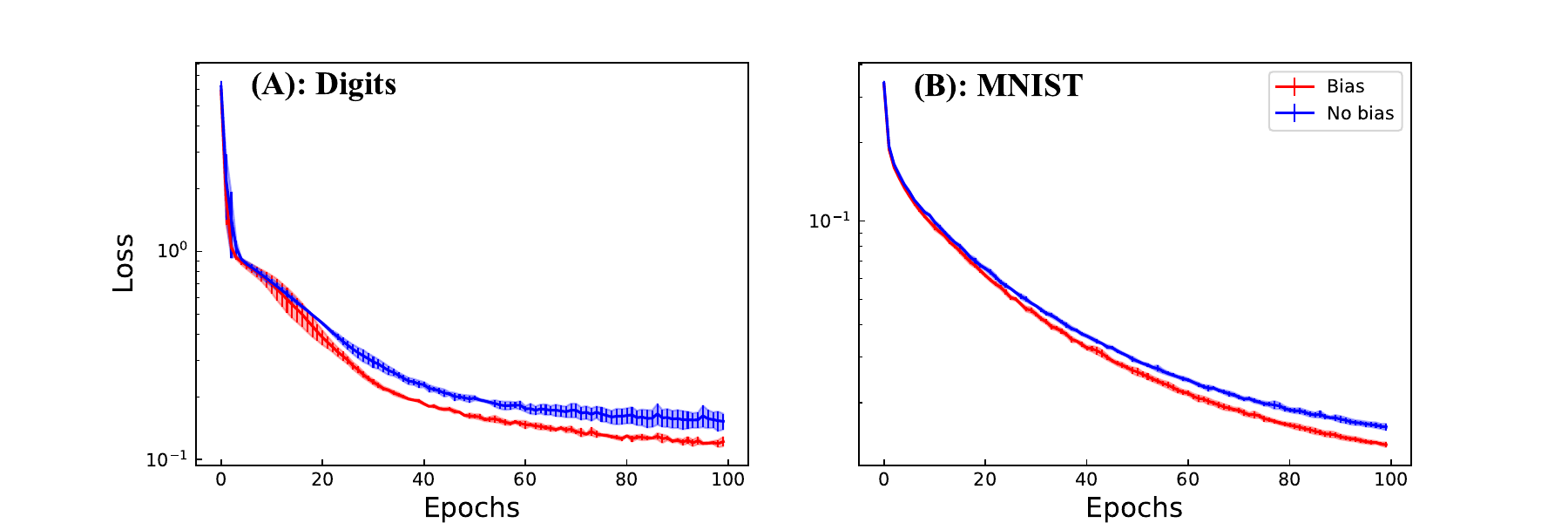}}

Figure \ref{f:Bias_impact_Digits_MNIST} shows the evolution of the training loss function plotted in logarithmic scale for 100 epochs for Digits database (left) and MNIST database (right). In both cases, we observe that models with bias slightly outperform models without bias. Moreover, for Digits, we also note that the standard deviation of the final obtained state is higher if there is no bias in the network, which means that training a network with bias increases its stability and reproducibility.

Therefore, it seems that the impact of an external drive is not very significant in large networks. Compared to fully-connected layers, bias represent around $0.2$-$0.3\%$ (for MNIST) of the total number of parameters. In smaller networks such as the one we trained on Digits, the gain in performance is a bit more significant, and the stability of training is increased. This could be interesting if we want to train a network in the presence of noise for example.

\end{document}